\documentclass[pdftex,twocolumn,epjc3]{svjour3}          

\RequirePackage[T1]{fontenc}

\smartqed  

\usepackage{amsmath}
\usepackage{bm}
\RequirePackage{graphicx}
\RequirePackage{mathptmx}      
\RequirePackage{flushend}
\usepackage{subfigure}

\usepackage{algorithm}
\usepackage{amssymb}
\usepackage[noend]{algpseudocode}
\algnewcommand\algorithmicforeach{\textbf{for each}}
\algdef{S}[FOR]{ForEach}[1]{\algorithmicforeach\ #1\ \algorithmicdo}

\usepackage{algorithm}
\RequirePackage[numbers,sort&compress]{natbib}
\RequirePackage[colorlinks,citecolor=blue,urlcolor=blue,linkcolor=blue]{hyperref}
\usepackage{nomencl}
\usepackage{siunitx}
\usepackage{todonotes}
\usepackage{amssymb}
\usepackage{mathtools}
 \usepackage{orcidlink}

\DeclarePairedDelimiterX{\infdivx}[2]{(}{)}{%
  #1\;\delimsize\|\;#2%
}

\DeclarePairedDelimiter{\norm}{\lVert}{\rVert}
\DeclareMathOperator\arctanh{arctanh}

\journalname{Eur. Phys. J. C}

\makenomenclature

\begin{document}
\title{End-to-end multi-particle reconstruction in high occupancy imaging calorimeters with graph neural networks}

\author{Shah Rukh Qasim\thanksref{cern,mmu}\orcidlink{0000-0003-4264-9724}
        \and
        Nadezda Chernyavskaya\thanksref{cern}\orcidlink{0000-0002-2264-2229}
        \and
        Jan Kieseler\thanksref{cern}\orcidlink{0000-0003-1644-7678}
        \and
        Kenneth Long\thanksref{mit}\orcidlink{0000-0003-0664-1653}
        \and
        Oleksandr Viazlo\thanksref{fsu}\orcidlink{0000-0002-2957-0301}
        \and
        Maurizio Pierini\thanksref{cern}\orcidlink{0000-0003-1939-4268}
        \and
        Raheel Nawaz\thanksref{sfu}\orcidlink{0000-0001-9588-0052}
}


\institute{Experimental Physics Department, CERN\label{cern}
          \and
          Manchester Metropolitan University\label{mmu}
          \and
          Florida State University\label{fsu}
          \and
          Massachusetts Institute of Technology\label{mit}
          \and
          Staffordshire University\label{sfu}
}

\date{Received: date / Accepted: date}

\maketitle

\begin{abstract}
\sloppy We present an end-to-end reconstruction algorithm to build particle candidates from detector hits in next-generation granular calorimeters similar to that foreseen for the high-luminosity upgrade of the CMS detector. The algorithm exploits a distance-weighted graph neural network, trained with object condensation, a graph segmentation technique. Through a single-shot approach, the reconstruction task is paired with energy regression. We describe the reconstruction performance in terms of efficiency as well as in terms of energy resolution. In addition, we show the jet reconstruction performance of our method and discuss its inference computational cost. To our knowledge, this work is the first-ever example of single-shot calorimetric reconstruction of ${\cal O}(1000)$ particles in high-luminosity conditions with 200 pileup.
\end{abstract}

\section{Introduction}
\sloppy The high-luminosity upgrade of the Large Hadron Collider (HL-LHC) will present unprecedented computing challenges~\cite{HEPSoftwareFoundation:2017ggl}. Because the processing complexity of LHC collision scales with the number of hits and energy deposits in the detectors from interacting particles, the computing resource needs will increase significantly as a function of the number of simultaneous proton-pair collisions at each particle beam crossing (pileup). In addition, particle detectors with irregular geometries, motivated by a need to combat the high radiation environment of high-pileup events, prohibit algorithms that view the detector as a simple grid and encourage more sophisticated approaches. 

One of the most problematic computing tasks is the so-called local event reconstruction, i.e., the task of clustering detector hits (energy deposits left by particles on various detector sensors) into particle candidates. Similar to how semantic segmentation works in computing vision, particle reconstruction is more than a clustering task. The reconstruction of a clustered object implies associating it to one particle category (a classification task) and determining the particle energy and flight direction (a regression task). In the future, the enforcement of 5D reconstruction (3D position, energy, and particle time of arrival) both in hardware and software would imply an additional regression task for the time to be considered.

In traditional approaches, particle reconstruction follows a two-step strategy: first, clusters are built, and then classification and regression tasks are performed on these clusters. In some cases, the regression and classification steps employ machine learning algorithms. Boosted decision trees have been used most extensively for this purpose at the LHC.
An example of this strategy is discussed in Ref.~\cite{CMS:2020uim}, in the case of the CMS electromagnetic calorimeter.

In LHC proton-proton collisions, two bunches of $\sim 10^{11}$ protons are brought to collision; to increase the probability of rare and interesting interactions (e.g., the production of a Higgs boson) to occur. Because of this, a single collision event contains the particles resulting from more than one collision (the primary particles). These particles travel through the detector components and, when crossing a calorimeter, produce showers of other particles (secondary particles). During Run II, an average LHC collision event consisted of $\sim 40$ pileup collisions, resulting in ${\cal O}(1000)$ primary particles. At the HL-LHC, up to 200 pileup collisions per event are expected. Traditional event reconstruction algorithms are challenged by the large primary-particle multiplicity due to the fast scaling of their memory utilization and execution time with the cardinality of the problem. Limiting the increase of computing resource consumption at large pileup is a necessary step for the success of the HL-LHC physics program~\cite{HEPSoftwareFoundation:2017ggl}. The use of modern machine learning techniques to perform particle reconstruction has been advocated as a possible solution to this problem~\cite{Albertsson:2018maf}.

In this paper, we present an end-to-end reconstruction algorithm, which takes as input a collection of detector hits in a highly granular calorimeter, similar to that under construction by the CMS collaboration~\cite{CMS:2017jpq}, in view of the HL-LHC upgrade~\cite{ZurbanoFernandez:2020cco}. The algorithm returns identified particle candidates and their momenta. The algorithm consists of a distance-weighted graph neural network~\cite{Qasim:2019otl,Iiyama:2020wap}, trained using object condensation~\cite{Kieseler:2020wcq}. At training time, the loss function is built by combining the clustering and regression tasks. In this respect, the presented algorithm is an example of a single-shot graph segmentation model that could find applications beyond the domain of particle physics. 

We discuss the algorithm performance in terms of accuracy and computational costs, considering in particular how its performance scales with increasing pileup. We consider both single-particle response and jet reconstruction to evaluate the reconstruction algorithm both at particle level and with higher-level objects. To our knowledge, this study represents the first demonstration of end-to-end calorimeter reconstruction at high pileup exploiting neural networks.

This paper is organized as follows: section~\ref{sec:related} discusses the existing literature related to this work; section~\ref{sec:detector} describes the detector geometry and the data generation workflow. Section~\ref{sec:data} describes the dataset format and its preprocessing. Sections~\ref{sec:model}~and~\ref{sec:opt} describe the model architecture and the optimization metric, respectively; section~\ref{sec:inference} discusses the inference clustering algorithm and how it improves with respect to the original version of object condensation and the computational costs; section~\ref{sec:physics} shows the physics performance of the algorithm, respectively; conclusions are given in section~\ref{sec:conclusion}.
\section{Related work}
\label{sec:related}
In recent years, GPU acceleration has been investigated as a means to speed up traditional particle-reconstruction algorithms by parallelization~\cite{Pantaleo:2016ery,Rohr:2017ydv,Rovere:2020rqi,Aaij:2019zbu,Bocci:2020olh,LHCb:2021kxm,Rohr:2021psv}. In view of these promising results, the LHC experimental collaborations invested financial resources to migrate their computing infrastructure towards CPU+GPU heterogeneous computing~\cite{LHCbCollaboration:2717938,Buncic:2011297,Collaboration:2759072}. This transition offers the possibility to exploit neural networks for the same task while benefiting from the impressive technological development in Artificial Intelligence (AI) inference on GPU+CPU heterogeneous platforms. Leveraging this technology trend, we are investigating the possibility of using solutions entirely based on AI algorithms to accomplish calorimetric reconstruction tasks~\cite{Qasim:2019otl,bhattacharya2022gnn,qasim2021multi}. A similar effort has already been established for particle tracking~\cite{Ju:2020xty,Choma:2020cry,Ju:2021ayy,DeZoort:2021rbj}. 

Neural networks have also been used in classification and regression tasks on portions of calorimeters~\cite{DeOliveira:2018szb,Belayneh:2019vyx,akchurin2021use}.
The use of neural networks for end-to-end reconstruction goes beyond calorimeter reconstruction or tracking.  Computer vision techniques based on convolutional and graph neural networks have been used for event-topology classification directly from the energy map of detector hits~\cite{Andrews:2018nwy,Andrews:2019faz,Andrews:2021mgj,thais2022graph}. Deep neural networks have been exploited as a tool to cluster high-level objects, e.g., particle-flow candidates~\cite{Pata:2021oez} and jets~\cite{Pol:2021iqw,Pol:2022ogc}. All these studies demonstrate that modern AI algorithms give significant accuracy and computational efficiency advantages, maximally benefiting from highly parallelizable hardware. For this reason, and given the popularity of these techniques outside the domain of particle physics, it is natural to assume that AI algorithms will be at the center of future technological developments and, consequently, will play a crucial role in the evolution of the computing model at future collider experiments.

\section{Detector}
\label{sec:detector}
For this study, we use a dedicated simulation of a highly granular calorimeter based on GEANT4~\cite{agostinelli2003geant4}. The detector geometry is a simplified version of the CMS high-granularity calorimeter (HGCAL)~\cite{hgcal_tdr}, expected to be integrated into the CMS detector for HL-LHC. 

\begin{figure}[t!]
\includegraphics[width=.45\textwidth]{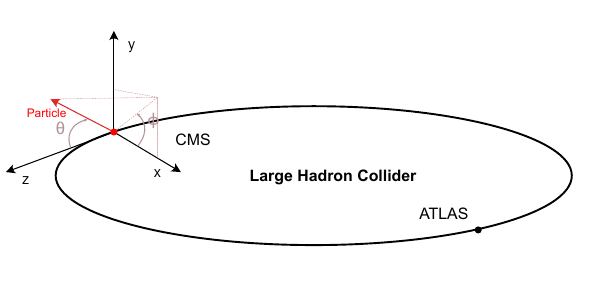}
\caption{Schematic representation of the right-handed Cartesian coordinate system adopted to describe the detector. \label{fig:ref_system}}
\end{figure}

To describe the detector, we use a right-handed Cartesian coordinate system with the $z$-axis oriented along the beam axis, the $x$-axis toward the center of the LHC, and the $y$-axis oriented upward, as shown in Fig.~\ref{fig:ref_system}. The $x$ and $y$ axes define the transverse plane, while the $z$-axis identifies the longitudinal direction. The azimuth angle $\phi$ is computed with respect to the $x$ axis. Its value is given in radians, in the [$-\pi, \pi$] range. The polar angle $\theta$ is measured from the positive $z$-axis and is used to compute the pseudorapidity $\eta = -\log(\tan(\theta/2))$. The transverse momentum ($p_T$) is the projection of the particle momentum on the ($x$, $y$) plane. We use natural units such that $c=\hslash=1$, and we express energy in units of electronvolt (eV) and its prefix multipliers.

\begin{figure*}[t!]
    \centering
    
   \includegraphics[width=.45\textwidth]{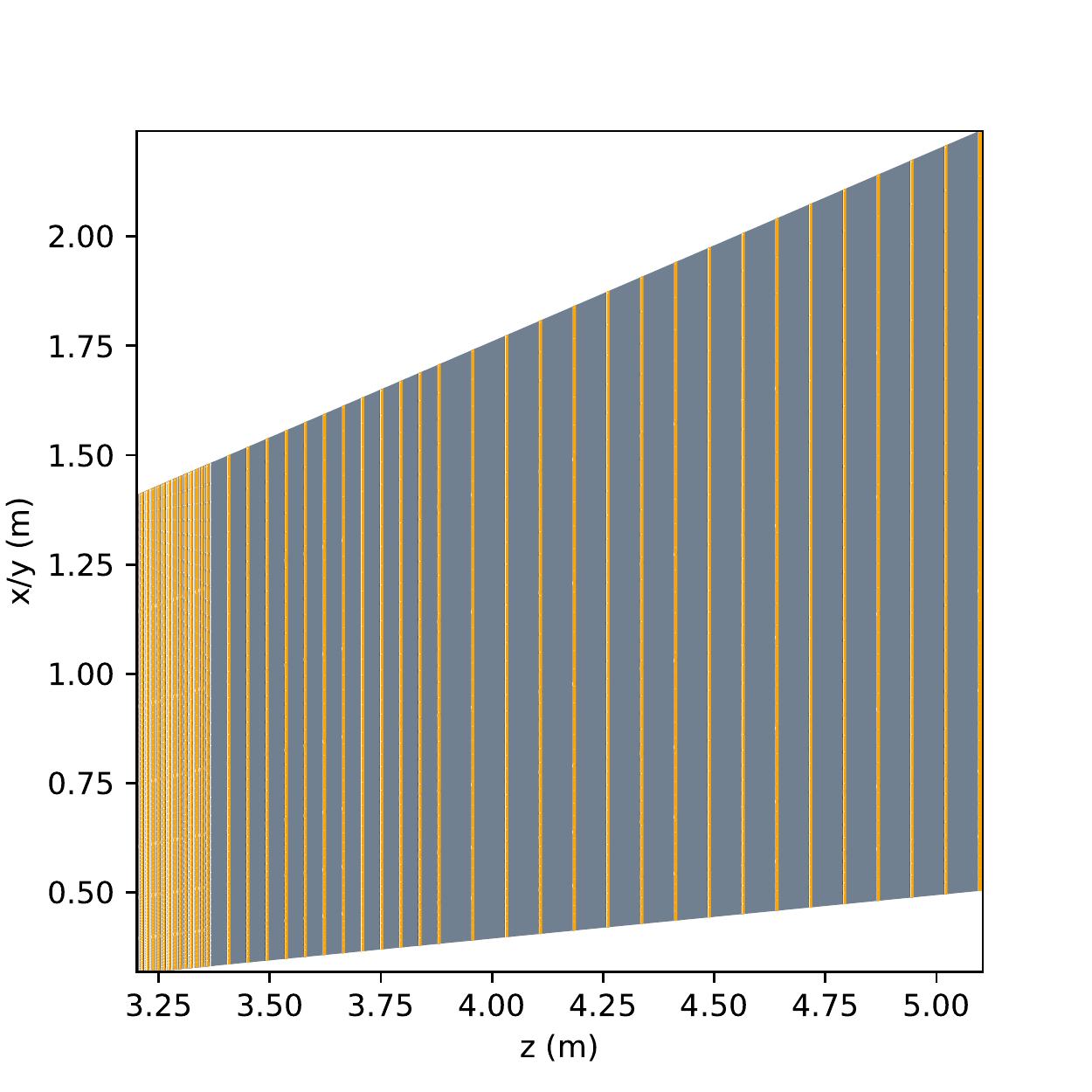}
   \includegraphics[width=.45\textwidth]{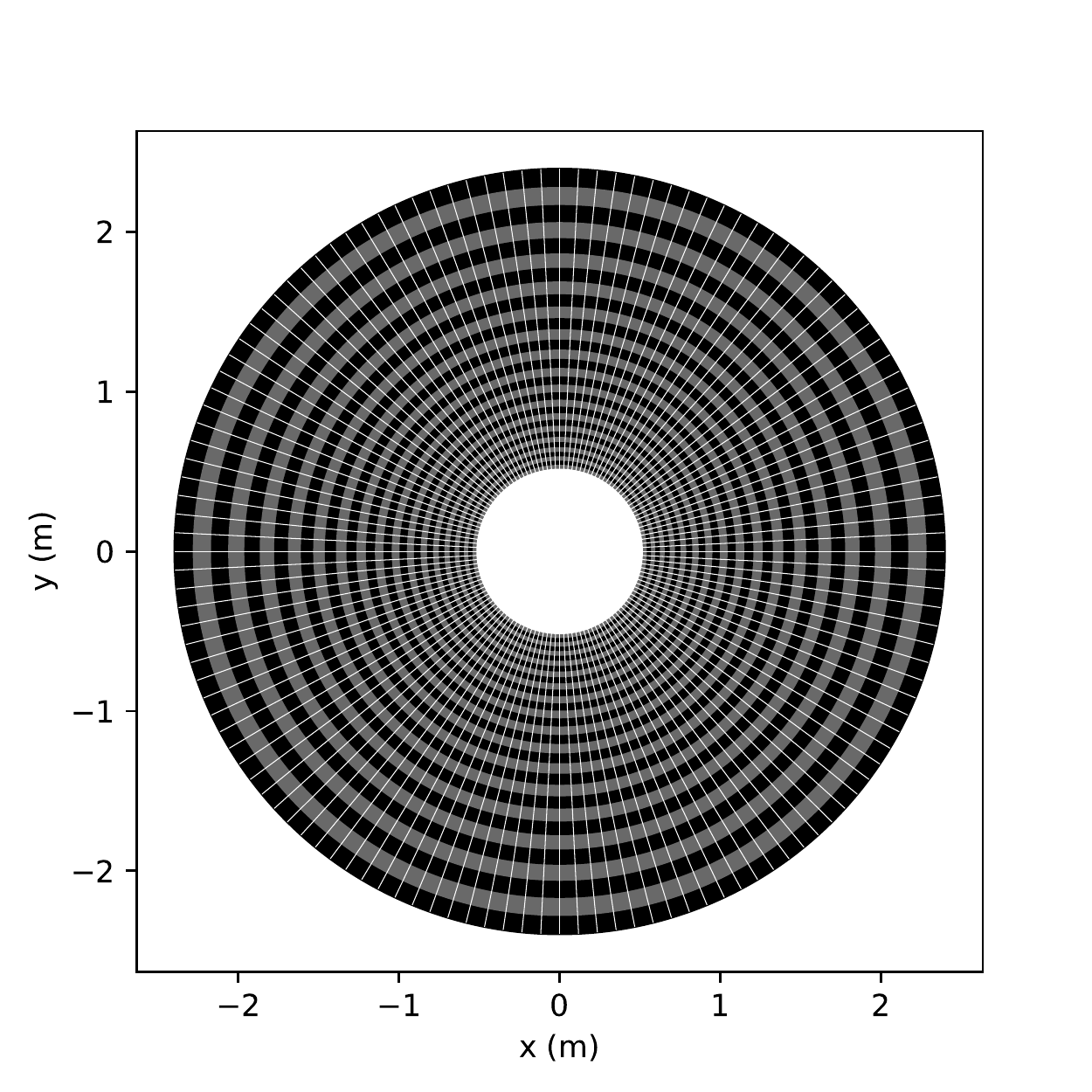}
    \caption{Left: Schematic representation of the detector longitudinal sampling structure. Right: Transverse view of the last active layer. Different colors represent different materials: copper (orange), stainless steel and lead (gray), air (white) and active sensors made of silicon (black).
    \label{fig:detector_cross_section}}
\end{figure*}

Like the CMS HGCAL, our detector is positioned in the endcap region of a cylinder-shaped $4\pi$ detector. The endcap calorimeter covers the $1.5<|\eta|<3.0$ pseudorapidity range. Its longitudinal and transverse cross-section views are shown in Fig.~\ref{fig:detector_cross_section}. The detector design is based on a sampling geometry structured in three blocks:
\begin{enumerate}
     \item 14x Electromagnetic layers
     \item 12x Hadronic layers with thin absorbers
     \item 16x Hadronic layers with thick absorbers
\end{enumerate}

\begin{figure}[t!]
    \centering
    \includegraphics[width=0.5\textwidth]{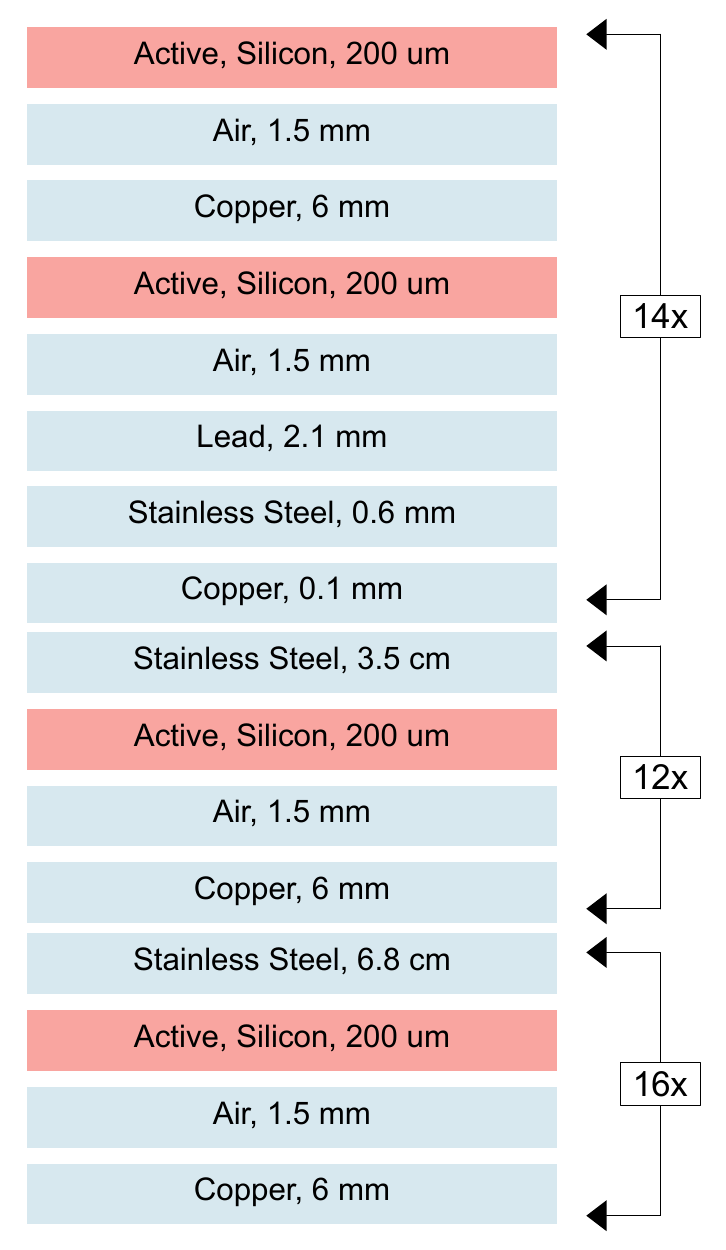}
    \caption{Layers of the detector. Top 8 blocks in the diagram represent the electromagnetic section of the calorimeter, and the rest, the hadronic section.}
    \label{fig:materials_diagram}%
\end{figure}

A schematic representation of the longitudinal sampling structure is shown in Fig.~\ref{fig:materials_diagram}, where the layer thickness in $z$ and the materials of which each each layer is composed are outlined. The electromagnetic section of the detector corresponds to 17 radiation and 1.3 nuclear interaction lengths, while the hadronic section corresponds to about 10 nuclear interaction lengths.

Since each electromagnetic layer consists of two silicon-sensor planes, there are, in total, 56 silicon layers. When projected on the ($\eta$,$\phi$) plane, the first silicon layer consists of square-shaped ~$0.02\times 0.02$ wide sensors. The sensor size linearly increases with the layer depth, reaching ~$0.07\times 0.07$ for the last layer. In total, each endcap consists of 778,712 sensors.
Due to cost-related considerations, the CMS HGCAL geometry is characterized by hexagon-shaped sensors. While being based on simpler square-shaped sensors, our detector has a comparable granularity and it is complex enough to faithfully represent the reconstruction challenges (e.g., in terms of image sparsity, average occupancy, and image resolution) of the HGCAL while making the GEANT4 simulation of the detector more tractable.

\section{Simulation and event generation}
\label{sec:data}
Data generation starts with a GEANT4 simulation of individual particles or individual proton collisions that are later combined to form more complex events, e.g., events with pileup. In each simulation, one or more primary particles are produced at the interaction point $(0,0,0)$. These particles travel in empty space, since no simulation of an inner tracking detector or magnetic field is part of our simplified setup. Some of the primary particles reach the calorimeter and interact with its material, starting a showering process that creates secondary particles $P_{\mathrm{secondary}}$. These secondary particles leave energy deposits on the sensors of the silicon layer as hits. Each hit on a sensor is associated to a secondary particle. 
Four types of interactions are simulated: 
\begin{enumerate}
    \item \sloppy \textbf{Type A}: Single-particle simulations for training. The particles are randomly chosen as $e^-$, $\gamma$, $\pi^{+}$, $\pi^{0}$ or $\tau$, with momentum coordinates uniformly distributed in $E \in [0.1, 200]$~GeV, $\eta \in [1.4, 3.1]$, and $\phi \in [0, 2\pi]$. $3.1\cdot10^5$ simulations are generated that are all used for training.
    \item \textbf{Type B}: Stable single-particle simulations generated 1 mm away from the detector, as if they are coming from the interaction point in a straight line for testing performance of the models. The particles are randomly chosen as $e^-$, $\gamma$, or $\pi^{+}$, with momentum coordinates uniformly distributed in $E \in [0.1, 200]$~GeV, $\eta \in [1.6, 2.9]$, and $\phi \in [0, 2\pi]$. $80,000$ simulations are generated which are all used for testing.
    \item \textbf{Minimum Bias}: Synthetic minimum bias proton-proton interactions, generated at a center-of-mass energy of $\sqrt{s}=13$~TeV (as during Run II of the LHC), using PYTHIA8~\cite{Sjostrand:2014zea}. $3.1\cdot10^5$ and $2\cdot10^5$ simulations are generated for training and testing, respectively. These simulations are used to generate pileup for both training and testing.
    \item \textbf{t$\bar{\text{t}}$}: Synthetic $q\overline{q} \to t\overline{t}$ events generated at $\sqrt{s}=13$~TeV using PYTHIA8. This sample is used to study the jet reconstruction accuracy. The $q \bar q$ production mechanism is selected in order to maximize the fraction of events that produce primary particles in the endcap region. $40,000$ simulations are generated, which are all used for testing.
\end{enumerate}

\begin{figure}[t!]
    \centering
    \includegraphics[width=0.5\textwidth]{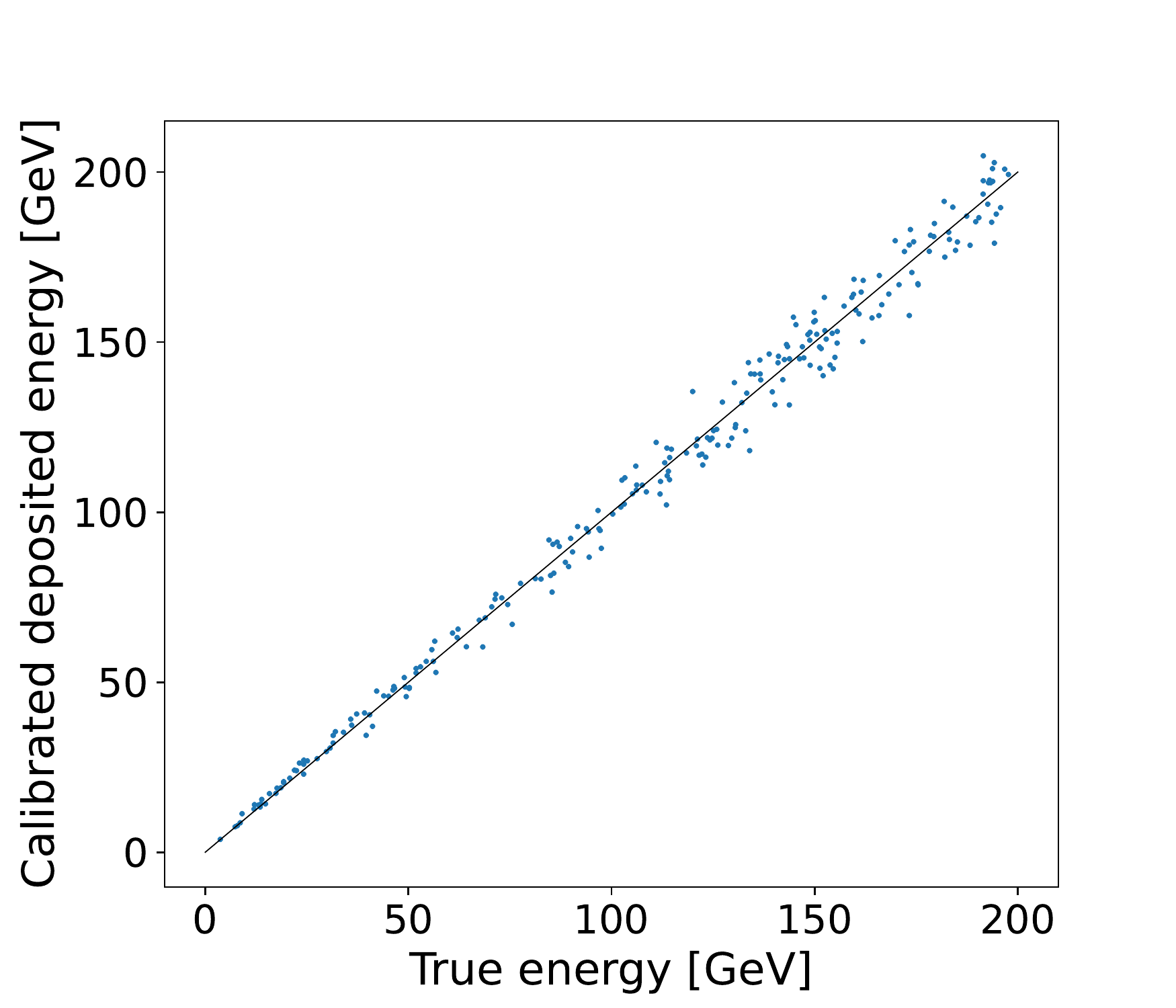}
    \caption{A scatter plot showing calibrated deposited energy versus true energy of photons shot between 1.8 $<$ $\eta$ $<$ 2.8.} 
    \label{fig:calibration_scatter}%
\end{figure}

As a pre-processing step, the raw energy deposit on a sensor returned by GEANT4 is calibrated by rescaling the hit energy according to a multiplicative factor $c_i$, defined as:
\begin{equation}
    c_i = 1 + \frac{w_a(i)}{w_s(i)}~,
\end{equation}
where $w_s(i)$ is the width in mm of the relevant sensor and $w_a(i)$ is that of the absorber layer in front of it. Starting from $c_i$, we then apply a global calibration factor $\hat{g}$, which is
computed by minimizing the squared difference between the energy of the incoming particle and the deposited hit energy, using a single-particle calibration dataset $D_{\mathrm{calib}}$ :
\begin{equation}
    \hat{g} = \underset{g}{\mathrm{argmin}}\sum_{S \in D_{\mathrm{calib}}}{(E(p_S)-\sum_{i \in \mathbb{S}}^{} gc_i~ d_{\mathrm{raw}}(i,p_S))^2}~~,
\end{equation}
where $p_S$ labels the unique primary particle in the event $S$, $E(p_S)$ - its energy, $i$ - the sensor in the ensemble $\mathbb{S}$ of all sensors, and $d_\mathrm{raw}(i,p_S)$ - the raw energy deposit on $i$ by $p_S$. The $D_{\mathrm{calib}}$ sample, a subsample of the training dataset, is defined by requiring all particles to be photons with $1.8 < \eta(p_{S}) < 2.8$, and consists of the 240 single-photon events with energies between $4$ and $200$ GeV. Fig.~\ref{fig:calibration_scatter} shows the calibration performance.

\begin{figure*}[h!]
    \centering
    \includegraphics[width=0.45\textwidth]{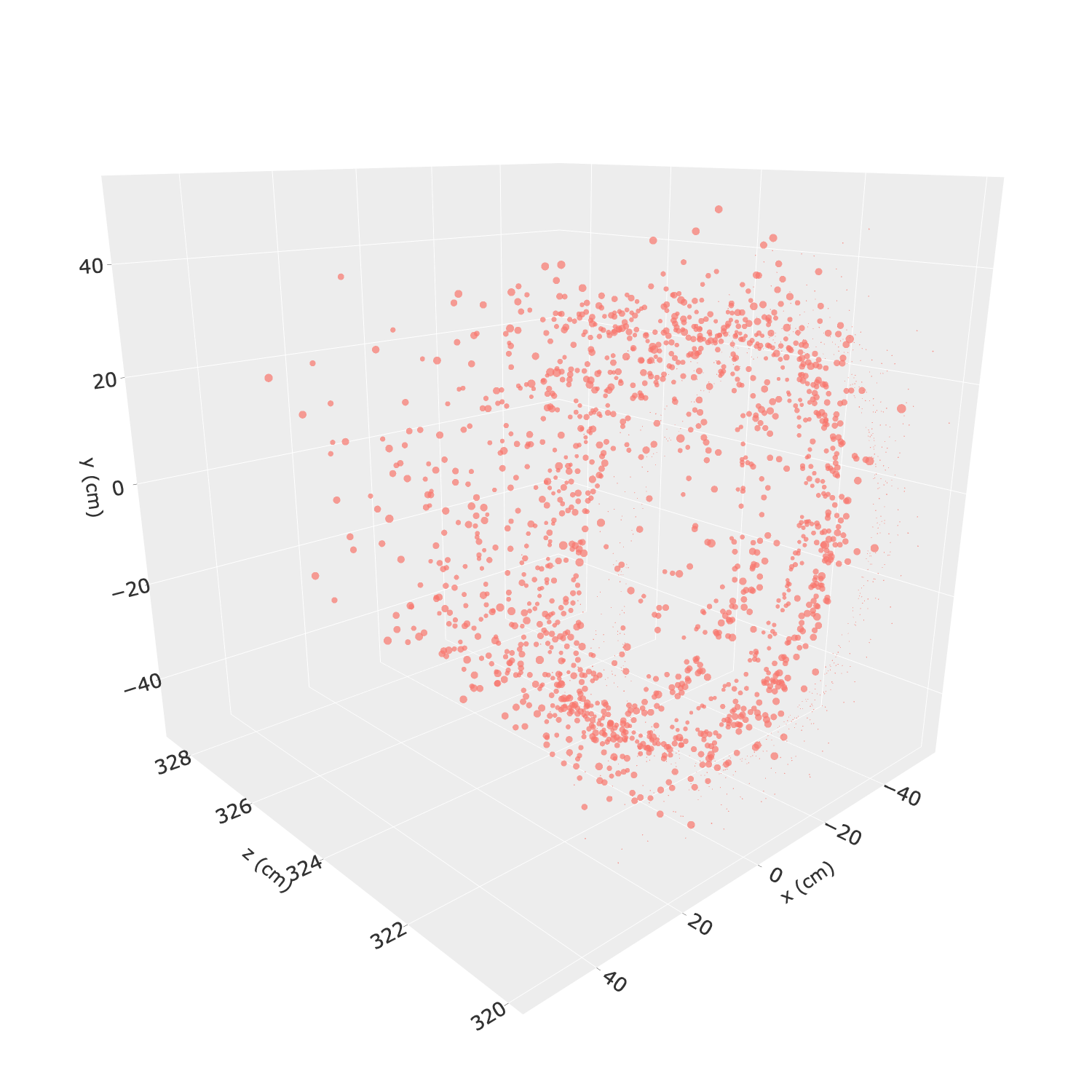}
    \includegraphics[width=0.45\textwidth]{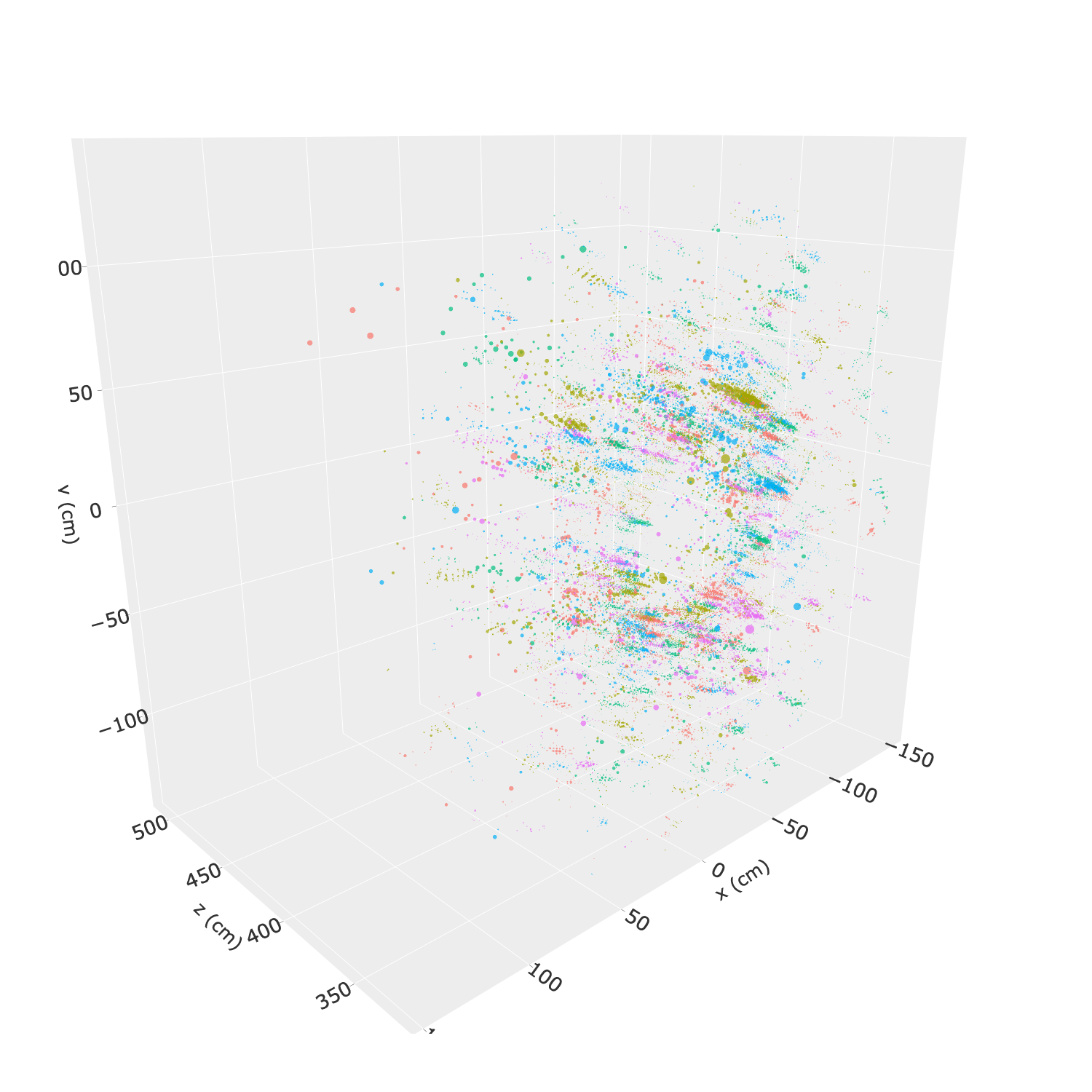}
    \caption[Left: Example of detector noise generated as described in the text. Right: Example of a generated event, obtained from the overlap of 40 single-collision events and noise. The shower of secondary particles generated by an individual primary particle yields a set of energy deposits in the detector labeled through markers of different colors and sizes. Different colors refer to different incoming primary particles. Marker sizes are proportional to $\log(e_h+1)$ and are max normalized independently in both figures.]{Left: Example of detector noise generated as described in the text. Right: Example of a generated event, obtained from the overlap of 40 single-collision events and noise. The shower of secondary particles generated by an individual primary particle yields a set of energy deposits in the detector labeled through markers of different colors and sizes. Different colors refer to different incoming primary particles. Marker sizes are proportional to $\log(e_h+1)$ and are max normalized independently in both figures.\footnotemark}
    
\label{fig:gen_event}
\end{figure*}

Several simulations ($\mathbf{S}$) are combined to form a full event as outlined in the algorithm Procedure~\ref{alg:event_gen}. The ultimate task on this full event is to reconstruct every particle that enters the detector. For this purpose, the ground-truth clusters ($T$) that the network is trained to learn are built from all the secondary particles ($P_{\mathrm{secondary}}$) from the combined simulations and not only from the primary ones. The raw deposits on the sensors from different simulations are added together. In order to emulate realistic detector conditions, detector noise is added to the deposited energy at generator-level, as specified in Procedure~\ref{alg:event_gen}:\ref{op:noise}. The detector noise model consists of a generation of spurious energy measurements in the detector sensors, distributed according to a Gaussian probability density function centered at 0 and with a variance of $5\cdot 10^{-6}$. An example of detector noise is shown in Fig.~\ref{fig:gen_event} (left).

\begin{algorithm}[h!]
    \floatname{algorithm}{Procedure}
    \label{alg:algo_generate}
    \caption{Event Generation}\label{alg:event_gen}
    \hspace*{\algorithmicindent} \textbf{Input} $\mathbf{S}$\\
    \hspace*{\algorithmicindent} \textbf{Output} $H, T$
    \begin{algorithmic}[1]
    
    \State $T \gets \bigcup_{\mathbb{S} \in \mathbf{S}} P_{\mathrm{secondary}}(\mathbf{S})$  \label{op:first}
    \State $G_\mathrm{close} \gets (T, \{f_{\mathrm{close}}(p_1, p_2) \forall (p_1,p_2) \in T\times T\})\footnotemark \label{op:f_close}$
    \State $T \gets \mathrm{merge}\left(\mathrm{connected\_components}\left(G_{\mathrm{close}}\right)\right)$
    
    \ForEach {$i \in \mathbb{S} $}
        \State $t(i) \gets \mathrm{undef}$
        \State $\overline{e} \gets \mathrm{sample}(N(0, 5 \times 10^{-6}))$ \label{op:noise}
        \State $e_{\mathrm{max}} \gets \mathrm{max}_T(d_{\mathrm{raw}}(i, p) \forall p \in T)$
        \If {$e_{\mathrm{max}} > \overline{e}$}
            \State $t(i) \gets \mathrm{argmax}_T(d_{\mathrm{raw}}(i, p) \forall p \in T)$ \label{op:hit_assignment}
        \EndIf
        \State $\overline{e} \gets \overline{e} + \sum_{p \in T}^{} d_{\mathrm{raw}}(i, p)$
        \State $e_i \gets \hat{g}c_i\overline{e_i}$

    \EndFor
    
    \State $H \gets \{i \forall i \in \mathbb{S} \ni \overline{e}_i/A_{i} > \rho\}$ \label{op_selection_hits} \label{op:picking}
    
    \State $T \gets \{p \forall p \in T \ni |{h \forall h \in H \ni t(h)=p}| > 0\}$  \label{op_selection_truth} \label{op:picking_true}

    \end{algorithmic}
\end{algorithm}
\footnotetext{Defines a graph as (nodes, edges)}



All sensors with uncalibrated deposited energy per sensor area ($\overline{e}/A$) greater than $\rho$ are considered as the reconstructed hits (rechits) in the event (Procedure~\ref{alg:event_gen}:~\ref{op:picking}). We choose the constant $\rho=1.3\cdot 10^{-7}$~GeV/mm$^2$, which corresponds to an uncalibrated energy ranging from $5.5\cdot 10^{-3}$~MeV to $3$~MeV between the smallest and the largest sensor. A set of rechits $H$ in the event is given as an input to the network. It is represented as a feature vector $V_{\mathrm{feat}}(h) = [r, \eta, \phi, x, y, z, \mathrm{A}, e]_h$ defined $\forall h \in H$, where $r, \eta, \phi$ ($x, y, z$) are the boost-invariant cylindrical coordinates (Cartesian coordinates) of the sensor, $\mathrm{A}$ is its area, and $e$ is the deposited energy. Multiple particles can leave an energy deposit in a single sensor, however the particle that leaves the highest deposit is considered as the true candidate ($t(s)$) for that hit (Procedure~\ref{alg:event_gen}:\ref{op:hit_assignment}).


We define the ground truth as a realistic target for the reconstruction algorithm, applying a selection to the generator-level hits in order to take into account their overlap. For instance, when two particles ($p_1$ and $p_2$) are maximally overlapping, we merge them into a single particle in the ground truth since disentangling such two clusters would be an impractical and imprecise task. The $f_\mathrm{close}(p_1,p_2)$ in Procedure~\ref{alg:event_gen}:\ref{op:f_close} evaluates whether the two showers are maximally overlapping if the following three conditions are met: their incident angles are closer in $\eta$ than $1.5 w_{\eta}$, in $\phi$ than $1.5 w_{\phi}$, and if the difference in their showering angles is less than $0.09$. $w_{\eta}$ and $w_{\phi}$ represent the max of width in $\eta$ and $\phi$, respectively, of the first sensors that $p_1$ and $p_2$ hit. The showering angle is taken from GEANT4 by selecting the momentum direction from when the particles start showering.

The set of true particles (including merged particles) $T$ that have at least one hit assigned to them after the filtering process are taken as the reconstruction target (Procedure~\ref{alg:event_gen}:\ref{op:picking_true}).

As an example, the 3D view of a generated event is shown on Fig.~\ref{fig:gen_event} (right). Detector hits belonging to different incoming primary particles are shown with different colors. 

\begin{figure}[h!]
    \centering
    \includegraphics[width=0.45\textwidth]{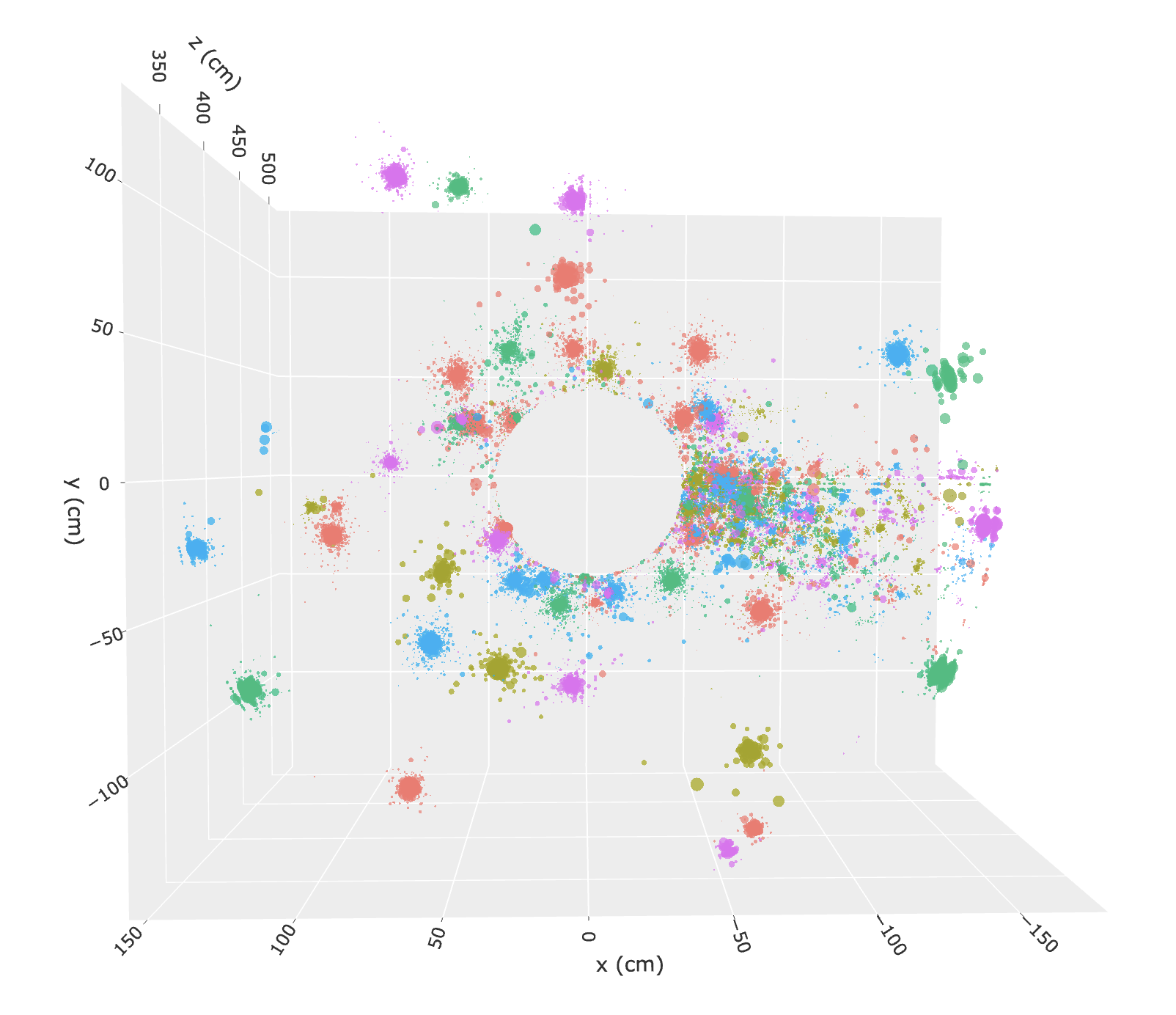}\label{fig:train_event_pu}
    \caption{Example of an event with pileup, part of the training set. No filter is applied to Type A simulations, but for pileup simulations, a filter, which selects particles (and their corresponding hits) between $\phi_0$ and $\phi_0+\ang{30}$, is applied. The high density of hits on the right side of the figure shows the slice in which pileup is present.\label{fig:train_event}}
\end{figure}

In total, six datasets are generated. Table~\ref{tab:event_complexity_x} shows the number of hits and true particles for different datasets as a proxy for event complexity.
\begin{enumerate}
    \item Training set: 5,000 events, where each event is created from 200 Minimum Bias simulations and 60 Type B simulations. However, due to computational constraints, it is impractical to train with 200 pileup. To overcome this, we augment the data. For each pileup simulation, we randomly picks a point in $\phi_0 \sim U(0, 2\pi)$ and only select $P_{\mathrm{secondary}}$ with impact directions between $\phi_0$ and $\phi_0+\ang{30}$. The particles originating from Type A simulations are left intact. Fig.~\ref{fig:train_event} shows an example event from the training set.
    \item Single-particle testing set: 20,000 events, where each event is created from Type B simulation only.
    \item PU40+1 testing set: 6,800 events, where each event is created from 40 Minimum Bias simulations and 1 Type B simulation.
   \item PU200+1 testing set: 6,800 events, where each event is created from 200 Minimum Bias simulations and 1 Type B simulation.
   \item PU40+$t\overline{t}$ testing set: 6,800 events, where each event is created from 40 Minimum Bias simulations and 1 $t\overline{t}$ simulation.
   \item PU200+$t\overline{t}$ testing set: 6,800 events, where each event is created from 200 Minimum Bias simulations and 1 $q\overline{q} \to t\overline{t}$ simulation.
\end{enumerate}

The training set with 5,000 events would require $10^6$ minimum bias simulations, but we could only generate $3.2\times10^5$. Therefore, we randomly sample from the simulation set without replacement. 
This strategy ensures minimum overlap of pileup between consecutive events and is used for generating all training and testing datasets.


\begin{table}[t]
    \caption{Event complexity for different datasets.}
    \begin{tabular}{l|l l l l}
     Dataset & $|H|$ & $|T|$ \\ \hline
     \textbf{Training set} & $34,000 \pm 2,000$ & $340 \pm 18$ \\ 
     \textbf{Single-particles testing set} & $2,600 \pm 240$ & $1.0 \pm 0.0$\\
     \textbf{PU40 testing sets} & $43,000 \pm 8,000$ & $1,000 \pm 160$ \\
     \textbf{PU200 testing sets} & $160,000 \pm 12,000$ & $3,200 \pm 130$ \\
    \end{tabular}
    \label{tab:event_complexity_x}
\end{table}

\section{Neural network and training}
\label{sec:model}

\begin{figure*}[t!]
    \centering
    \includegraphics[width=0.9\textwidth]{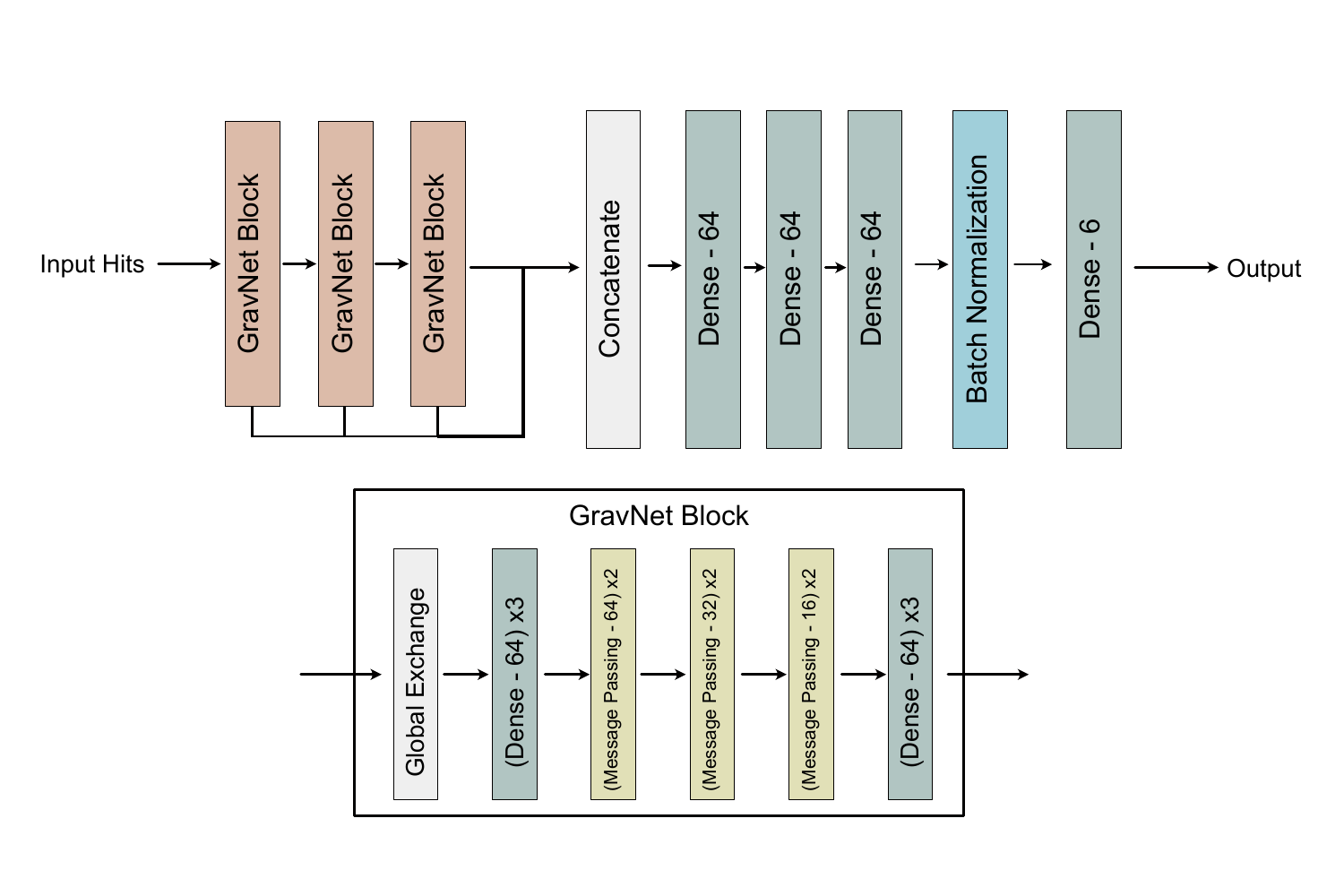}
    \caption{Architecture of the model. Three GravNet blocks are used, each of them containing multiple message passing layers. In a ResNet-type architecture, the features output from these layers are concatenated and then passed through a dense net. For each hit, six outputs are produced. Three represent the clustering space, one $\beta$ confidence, one distance threshold, and one energy correction factor.}
    \label{fig:architecture}%
\end{figure*}

As discussed in section~\ref{sec:data}, the input to the neural network is a set of hits. To create the edges needed to construct a graph, we use a dynamic graph neural network approach, GravNet~\cite{Qasim:2019otl}. GravNet dynamically computes the edges with the help of the k-nearest-neighbour algorithm, evaluated in a low dimensional learnable coordinate space. With each application of a GravNet layer, this coordinate space can change, and as a consequence, a different set of neighbours can be assigned to each vertex. Fig.~\ref{fig:architecture} visualises the model architecture, which is inspired by Ref.~\cite{qasim2021multi} with minor modifications.

Following the object condensation paradigm, the model makes a set of predictions per hit: the three-dimensional cluster space coordinates $x_h$, a condensation score $\beta_h$, and other object properties, in this case an energy correction factor $\psi_h \approx 1$. This factor corrects the total deposited calibrated energy in the calorimeter cells, evaluated using truth, to match the impact energy of the particle the hit belongs to. The total calibrated deposited energy left by a particle on all the sensors is defined as the deposited energy of the said particle. In addition, we introduce a distance measure $\varphi_h$, also per hit, that scales with the expected distance between one shower and hits from other showers or noise. This addition to object condensation allows us to keep the cluster coordinate space low dimensional, and therefore interpretable, while introducing another degree of freedom to adopt distances to locally dense environments. In total, the neural network has a 6-dimensional output per hit. These quantities are further explained in section~\ref{sec:opt} and section~\ref{sec:inference}.


\footnotetext{The Same method is used for choosing marker size for all the event displays in this article.}The model is trained for 68 epochs using the Adam optimizer~\cite{kingma2014adam} on the training set described in section~\ref{sec:data}, at which point the loss becomes stable and does not further decrease with additional epochs. Each training batch consists of one event\footnote{This choice is governed by the available GPU memory, not due to restrictions of the training framework.}. The training is performed within the DeepJetCore framework~\cite{DeepJetCore} and the models are implemented in Tensorflow~\cite{tensorflow} and Keras~\cite{keras}. We run the training at the Flatiron GPU computing cluster using NVIDIA V100 GPUs.




\section{Loss function}
\label{sec:opt}

The object condensation method~\cite{Kieseler:2020wcq} aims at identifying a unique representative hit accumulating all shower properties for each shower, referred to as condensation points. Each hit is also embedded in the clustering space to resolve ambiguities and to assign the remaining hits to showers. The loss consists of three terms: the potential loss ($L_V$) is responsible for creating the clustering space and embedding hits in it, the condensation score loss ($L_{\beta}$) trains the network to identify the condensation points, and the payload loss $L_{P}$ creates gradients for the other object properties, in our case the energy correction factor. The relative contribution of the two loss terms is set by a factor $s_C$, which we take as $s_C=1$ for this study. 

\begin{equation}
    L = L_V + s_C(L_{\beta} + L_{P})
\end{equation}

Especially for the hadronic showers, hits that are significantly displaced in position from the shower core are challenging to assign to their initiating incident particle. The mis-assignment of these hits to showers that are closer in space, but initiated by another incident particle is relatively common and difficult to avoid. Typically such hits are low energy, so their correct shower association is less critical to the task of estimating the total shower properties. With this in mind, we adapt the original object condensation method to reduce the impact of the mis-association of this class of hits in the network. Mathematically, this also serves to reduce the maximum fluctuations in the gradients. In comparison to Ref.~\cite{Kieseler:2020wcq}, we adapt the calculation of the clustering charge $q_h$, and smoothen the potential terms.
The clustering charge is calculated based on $\beta_h$ with $0\leq \beta_h \leq 1$. We rescale the calculation of $q_h$ slightly to avoid strong gradients for $\beta_h \to 1$ as follows:


\begin{equation}
    \label{eq:q}
    q_h = \arctanh^2(\beta_h/1.002) + q_{\mathrm{min}}\nu_h
\end{equation}

In addition, a new parameter $\nu_h$ that describes a spectator weight is introduced. Hits that are scattered far away from the shower core receive a smaller weight, in our case $\nu_h=0.1$, while all other hits receive a weight of $\nu_h=1$. To define whether a given hit should be considered a spectator, we first perform a principal component analysis (PCA) on the truth-assigned energy-weighted hit coordinates of the shower to identify the two principal components which act as the proxies for the shower axes. For this task we define the shower axes in two dimensions only, where the dimensionality is reduced by one due to the shower symmetry. The hits belonging to the shower are then projected onto the defined shower axes. Using the projected coordinates, we compute the Mahalanobis distance~\cite{mahalanobis1936generalized} for each hit. We consider a hit a spectator if its Mahalanobis distance is larger than 3.
For the attractive and repulsive potential losses, the hit $\alpha_t$ with the highest $\beta$ score for each truth shower $t$, also taking into account the spectator weights, plays a special role. It is defined as:

\begin{equation}
    \label{eq:alpha_assignment}
    \alpha_t = \underset{h \in H_t}{\mathrm{argmax}}(\beta_h \nu_h) \text{,}
\end{equation}

where $H_t$ is the set of hits belonging to truth shower $t$.
Furthermore, the $\beta$-weighted average learned distance scale $\overline{\varphi}_{t}$ for a truth shower $t$ is calculated as: 

\begin{equation}
\label{eq:phibar}
    \overline{\varphi}_{t} = \frac{\sum_{h \in H_t} \beta_h*\varphi_h}{\sum_{h \in H_t} \beta_h} \text{.}
\end{equation}

Taking the weighted average over the shower as opposed to considering only the hit $\alpha_t$ has the advantage that it creates a more consistent gradient for $\varphi_h$ to be learned for every hit.
A similar approach is taken for the reference point in clustering space of the potentials that attract or repulse other hits. Here, the reference point for each truth shower $t$ is calculated as

\begin{equation}
    \label{eq:xbar}
    \overline{x}(t) = \frac{1}{2}\left(x_{\alpha_t} + \frac{\sum_{h\in H_t}(\beta_h q_h x_h)}{\sum_{h\in H_t}{\beta_h q_h}} \right) \text{.}
\end{equation}
This represents another modification of the original object condensation loss, which takes $x_{\alpha_t}$ only. The new term in the sum serves to remove noise from the training, while keeping a large impact of the hit $\alpha_t$, which helps to resolve the degeneracy in the beginning of the training. Based on these ingredients, the attractive potential loss, $\breve{V}_h$, is then re-defined as follows:

\begin{equation}
    \label{eq:attract}
    \breve{V}_t(h) = q_{\alpha_t} w_{t} \ln\left(e\cdot\left( \frac{\norm{x_h-\overline{x}(t)}^2}{2\overline{\varphi}_{t}^2+\epsilon}\right) +1\right) \text{,}
\end{equation}
where $w_t$ is the shower weight. For $E_{\mathrm{true}} > 10$, $w_t=1$. From $10$ to $0.5$ GeV, it linearly decreases from $1$ to $0$. $\epsilon$ is a small number that is added for numerical stability. The repulsive loss is modified accordingly, as

\begin{equation}
    \label{eq:repulse}
    \hat{V}_t(h) = q_{\alpha_t}w_{t} \cdot \mathrm{exp}\left({- \frac{\norm{x_h-\overline{x}(t)}^2}{2\overline{\varphi}_{t}^2+\epsilon}}\right) \text{.}
\end{equation}
The full potential loss function takes the form:
\begin{equation}
    L_V = \frac{1}{|T|} \sum_{t \in T} \left( \frac{1}{|H_t|}\sum_{h \in H_t} q_h \breve{V}_t(h) + \frac{1}{|H-H_t|}\sum_{h \in (H-H_t)} q_h \hat{V}_t(h)  \right) \text{.}
\end{equation}
Here $H-H_t$ represents the set difference, i.e., all hits that are not assigned to shower $t$. The payload loss $L_P$ is also weighed by the object weight $w_t$ to reduce the impact of low energy showers, that is

\begin{equation}
    \label{eq:corr}
    L_P =  \sum_{t \in T}{} \frac{w_t}{\sum_{h \in H_t}^{} \xi(h)} \sum_{h \in H_t}^{} \xi(h) L_E  \text{,}
\end{equation}
with $ \xi(h) = \arctanh^2(\beta_h/1.002)$.
The energy loss contribution $L_E$ is calculated as
\begin{equation}
    L_E =\mathrm{log}\left( \left(\frac{E_{\mathrm{true},t} - \psi_h E_{\mathrm{dep},t}}{\sqrt{E_{\mathrm{true},t}} + 0.003}\right)^2+1\right) \text{,}
\end{equation}
where $E_{\mathrm{dep}}$ is the total energy collected in the calibrated calorimeter cells associated to the truth shower $t$. 


The beta loss term consists of two parts and is identical to Ref~\cite{Kieseler:2020wcq}, 
\begin{equation}
    L_{\beta} = \frac{1}{|T|}\sum_{t\in T} (1-\beta_{\alpha_t}) + s_B\frac{1}{|H_{\circ}|} \sum_{h \in H_{\circ}} \beta_h \text{.}
\end{equation}
The first term ensures that at least one hit per truth shower is promoted to a condensation point. The second term suppresses noise. $H_{\circ}$ represents the set of all noise hits. We choose the scaling factor $s_B = 1$.



\section{Inference}
\label{sec:inference}


We also extend the inference algorithm from Ref~\cite{Kieseler:2020wcq} to reflect the introduction of the local distance scale $\phi_h$. The algorithm is outlined in Procedure~\ref{alg:algo_inference_clustering} and is applied to the learned clustering space. The algorithm starts with the hit with the highest $\beta$-score $\beta_\alpha$ and assigns all hits within a certain radius $t_d \cdot \phi_\alpha$ to it, with $t_d=1.0$. These hits are removed for the next iteration.
This procedure is repeated until the highest $\beta$-score is lower than the threshold $t_\beta = 0.3$. The remaining unassigned hits are considered noise.
To determine the energy of the reconstructed cluster, we sum the energy of all hits assigned to a cluster collected around hit $\alpha$ and multiply this sum by the learned energy correction factor $\psi_\alpha$.

\begin{algorithm}[h]
    \floatname{algorithm}{Procedure}
    \caption{Clustering Inference}\label{alg:algo_inference_clustering}
    \hspace*{\algorithmicindent} \textbf{Input} $H, \beta, x, \psi, \varphi, t_d, t_{\beta}$\\
    \hspace*{\algorithmicindent} \textbf{Output} $P$
    \begin{algorithmic}[1]

    \State $\mathbf{P} \gets \{\}$
    
    \State $H_{\mathrm{cand}} \gets \{h \in H \forall h \in H \ni \beta_h > t_{\beta}\}$
    \State $H_{\mathrm{free}} \gets H$

    \While {$|H_{\mathrm{cand}}| > 0$}
        \State $\alpha \gets \mathrm{argmax}_h(\beta_h \forall h \in H_{\mathrm{cand}})$
        \State $p \gets \mathrm{NEW\_PARTICLE}$
        \State $H_p \gets \{h \in H_{\mathrm{free}}, \norm{x_h - x_{\alpha}} < t_d \varphi_ {\alpha}\}$ \label{op:local_radius_clustering}
        
        \State $E_{\mathrm{pred}}(p) \gets \psi_{\alpha}\sum_{h\in H_p} e_h$ \label{op:corr_assignment}
        
        \State $H_{\mathrm{free}} \gets H_{\mathrm{free}} - H_p$
        \State $H_{\mathrm{cand}} \gets H_{\mathrm{cand}} - H_p$
        
        \State $P \gets P \cup \{p\}$
        
    \EndWhile
    \end{algorithmic}
\end{algorithm}

\begin{figure}[t!]

\includegraphics[width=.45\textwidth]{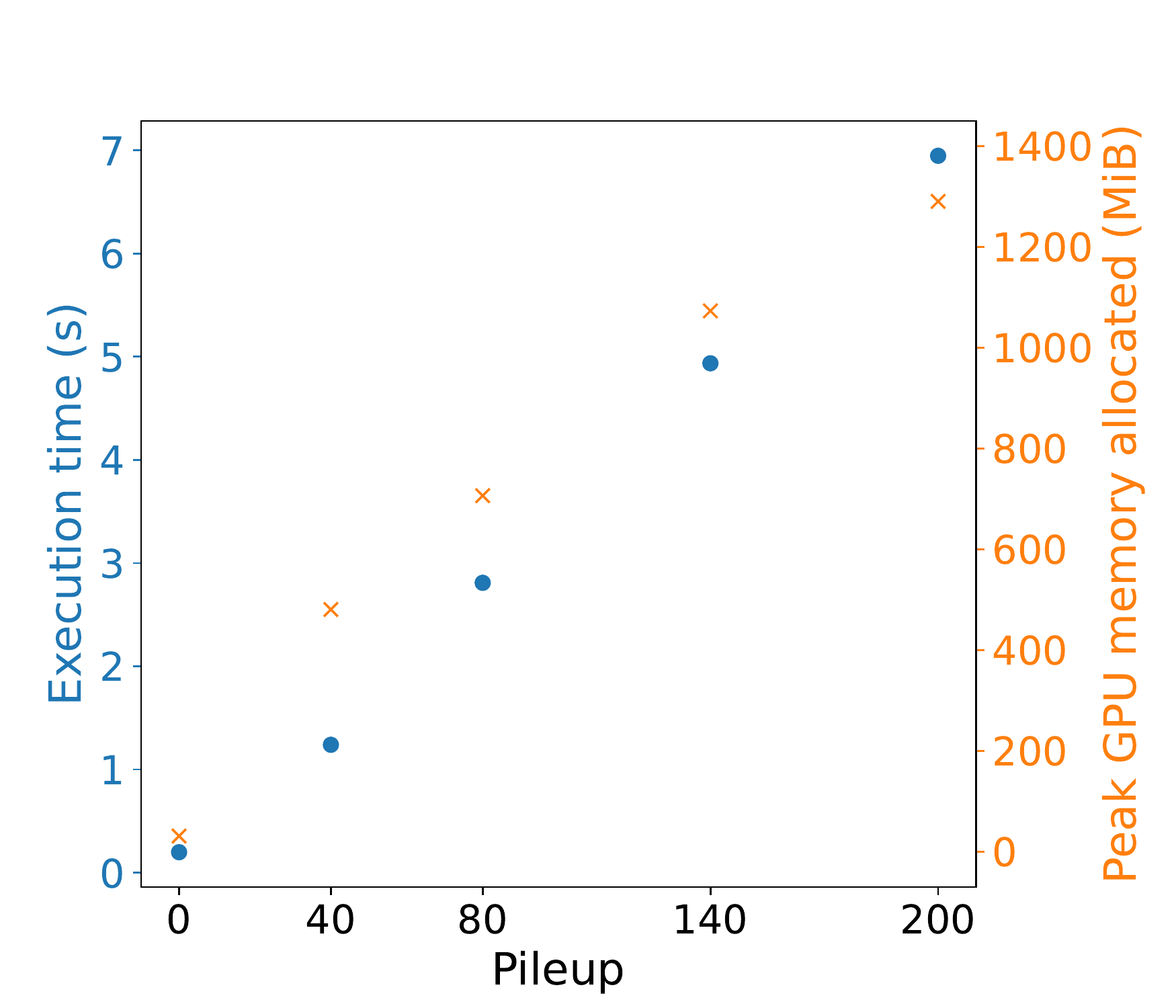}
\caption{Compute specifications of our model as a function of the amount of pileup to run inference with one event. The left axis (blue) shows average execution time, and the right (orange) shows the average peak memory allocated on the GPU.\label{fig:resource_req}}
\end{figure}

In Fig.~\ref{fig:resource_req}, we show inference time and peak GPU memory required for single particle in different pileup conditions. The inference times are evaluated on a Nvidia V100 GPU.  For single-particle events, inference takes about 200 ms. The inference time increases to 1.2 s or 7 s for 40 and 200 pileup, respectively. We expect an additional significant improvement of the overall inference time using edge-contraction methods to reduce the cardinality of the hits. These refinements will be considered in future work. These values should be compared to ${\cal O}(1000)$ s taken by currently adopted algorithms running on CPU, when scaled up to a 200 pileup environment.

Inference in 40 and 200 pileup allocates an average of only 500 MiB and 1300 MiB, respectively, on the GPU. This opens up the possibility that our method can be deployed on machines with less powerful GPUs with smaller VRAM. Note that a larger GPU is required for the training stage as memory can't be freed up after executing a neural network layer for backpropagation-related computations.

Here, the final inference algorithm was adapted to only consider close-by hits using a binning approach, making its contribution to the execution time negligible.

\section{Physics performance}
\label{sec:physics}
We evaluate physics performance in several ways by studying the reconstruction performance of the individual particles and jets. The individual particles, split in electromagnetic particles ($e^-$ and $\gamma$) and hadronic particles ($\pi^+$), are studied separately as they exhibit different behaviors. Reconstruction efficiency, energy response, and resolution are studied in different pileup environments, as well as the rate of reconstructed clusters that are either split off from the main shower (unmatched showers). For jets, we investigate the response and resolution in different pileup environments, assuming per-particle pileup removal procedures are in place.

The metrics are studied as a function of the $p_T$ of the particles and jets. The neural network is regressing only the particles' energy, but for the computation of their $p_T$, we use energy-weighted mean hit positions to estimate the particles' direction. For consistency, we also use the same methodology to compute truth-level $p_T$. 

Figure~\ref{fig:event_reco_examples} visually shows the predictions of the neural network and compares them to the truth for both individual particle reconstruction and jet reconstruction.

\subsection{Particle reconstruction performance}
We begin by studying the performance in 0 pileup. These events contain only one probe truth particle and some detector noise. The probe particle is taken from Type B simulations as discussed in section~\ref{sec:data}. We then overlay the probe particle with 40 and 200 pileup interactions to study performance in a more controlled fashion. While our method reconstructs all the particles in the event, including all the particles from the pileup, we only study the reconstruction performance of the probe particle.

\begin{figure*}[ht]
   \centering
   \subfigure[]{\includegraphics[width=.45\textwidth]{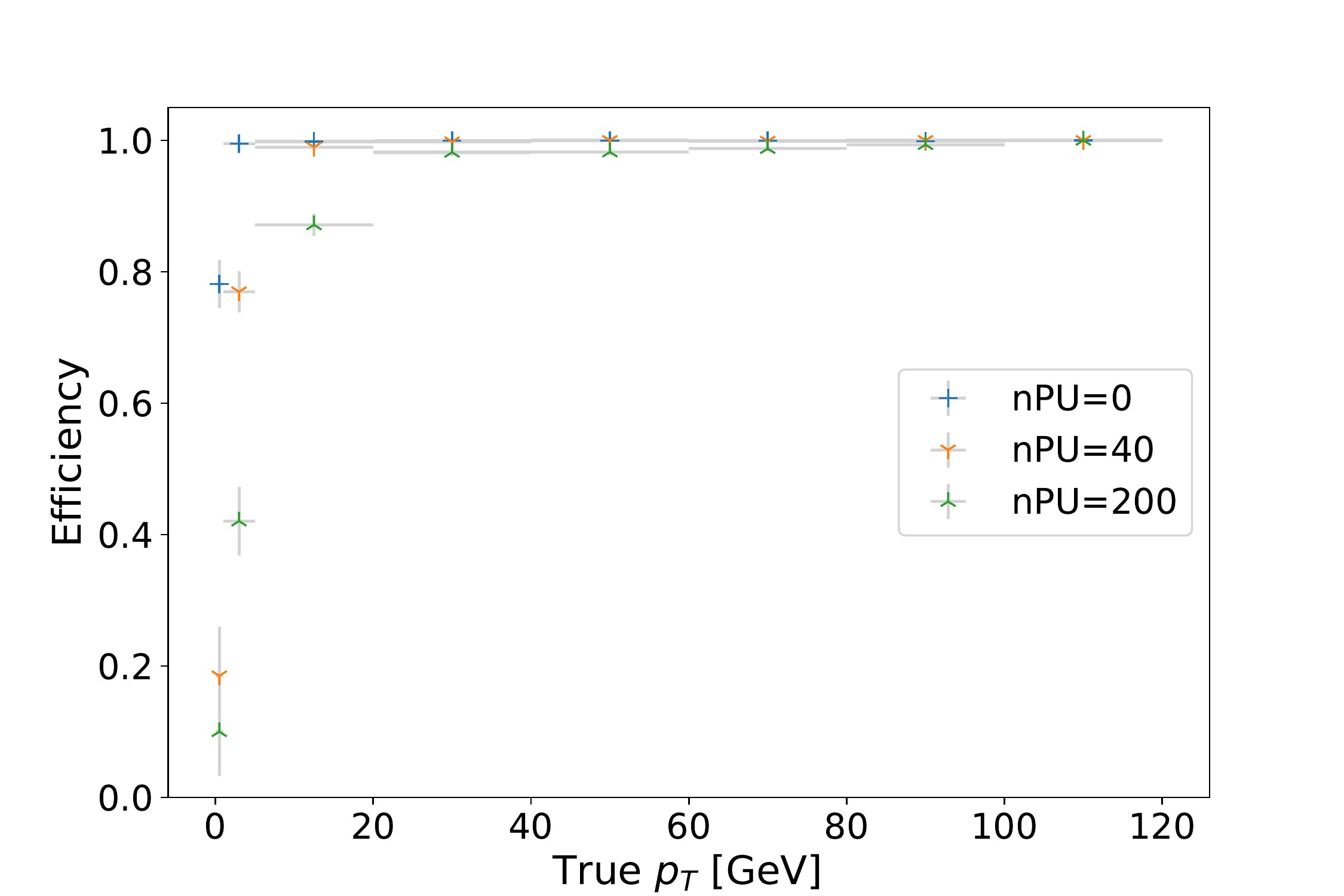}\label{fig:em_eff}}\quad
   \subfigure[]{\includegraphics[width=.45\textwidth]{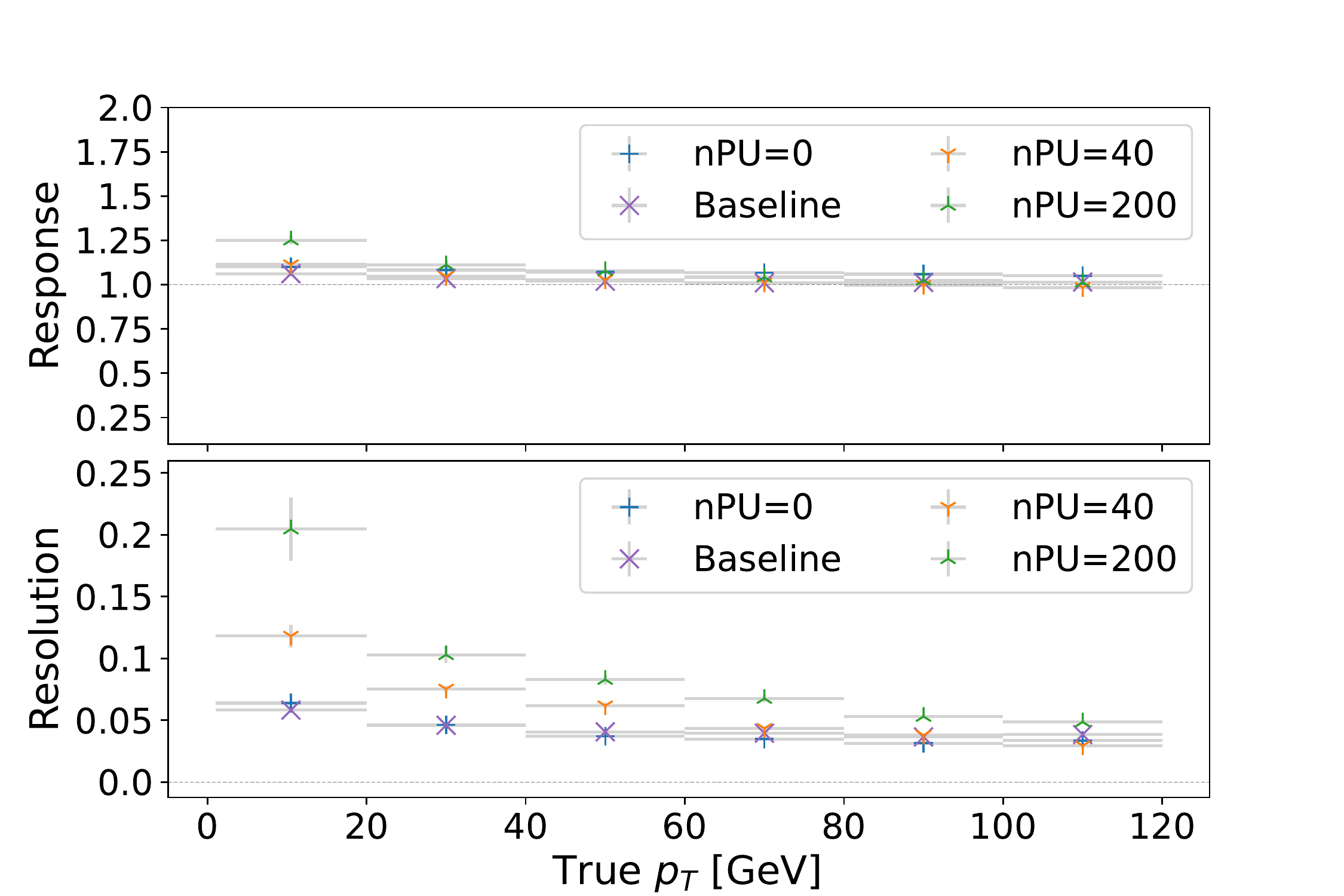}\label{fig:em_res}}\\
   \subfigure[]{\includegraphics[width=.55\textwidth]{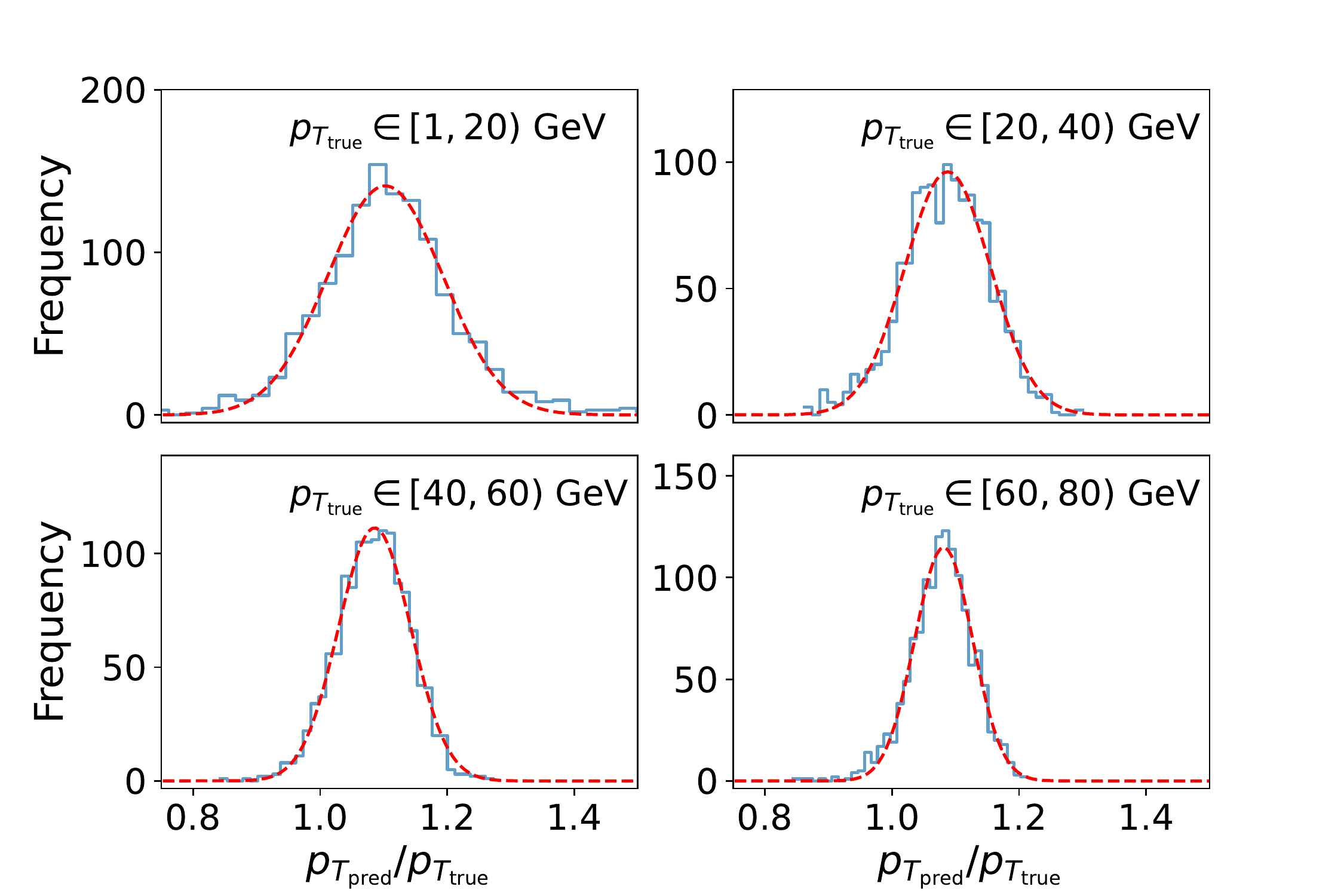}\label{fig:em_fit}}
   \caption{Reconstruction performance of electromagnetic particles (photons and electrons) in different pileup environments. Fig.~\ref{fig:em_eff}: Efficiency as a function of the true $p_T$. Fig.~\ref{fig:em_res}: Mean response (top) and resolution (bottom) as a function of the true $p_T$. The response and resolution are computed as the mean and mean-corrected standard deviation of the Gaussian fit to the ${{p_T}_{\mathrm{pred}}}/{{p_T}_{\mathrm{true}}}$ distribution in individual $p_T$ bins. Fig.~\ref{fig:em_fit}: The distribution of $p_T$ response in different $p_T$ ranges corresponding to the first four bins in Fig.~\ref{fig:em_eff} and Fig.~\ref{fig:em_res} in 0 pileup environment.}
   \label{fig:em_results}
\end{figure*}

\begin{figure*}[ht]
   \centering
   \subfigure[]{\includegraphics[width=.45\textwidth]{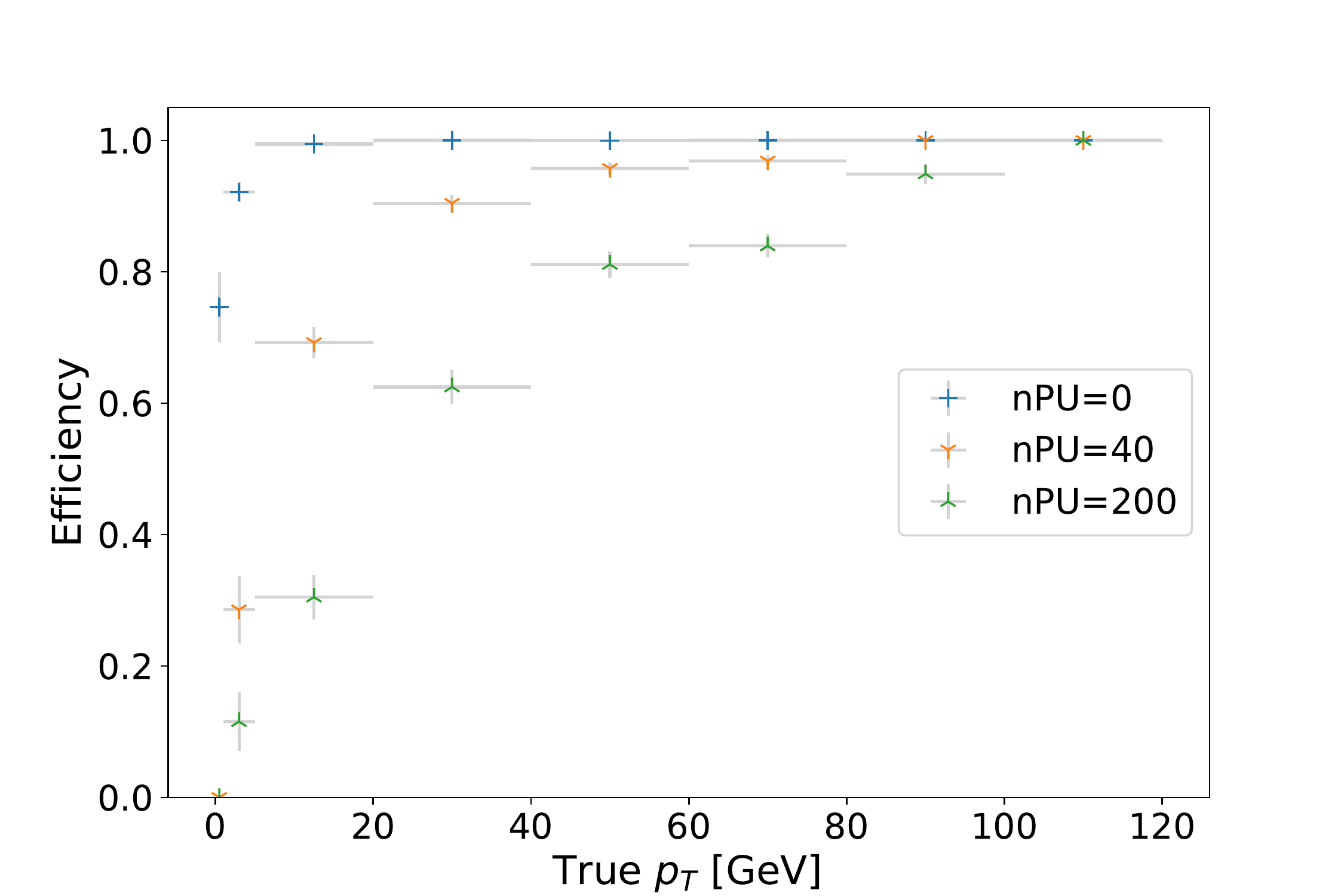}\label{fig:had_eff}}\quad
   \subfigure[]{\includegraphics[width=.45\textwidth]{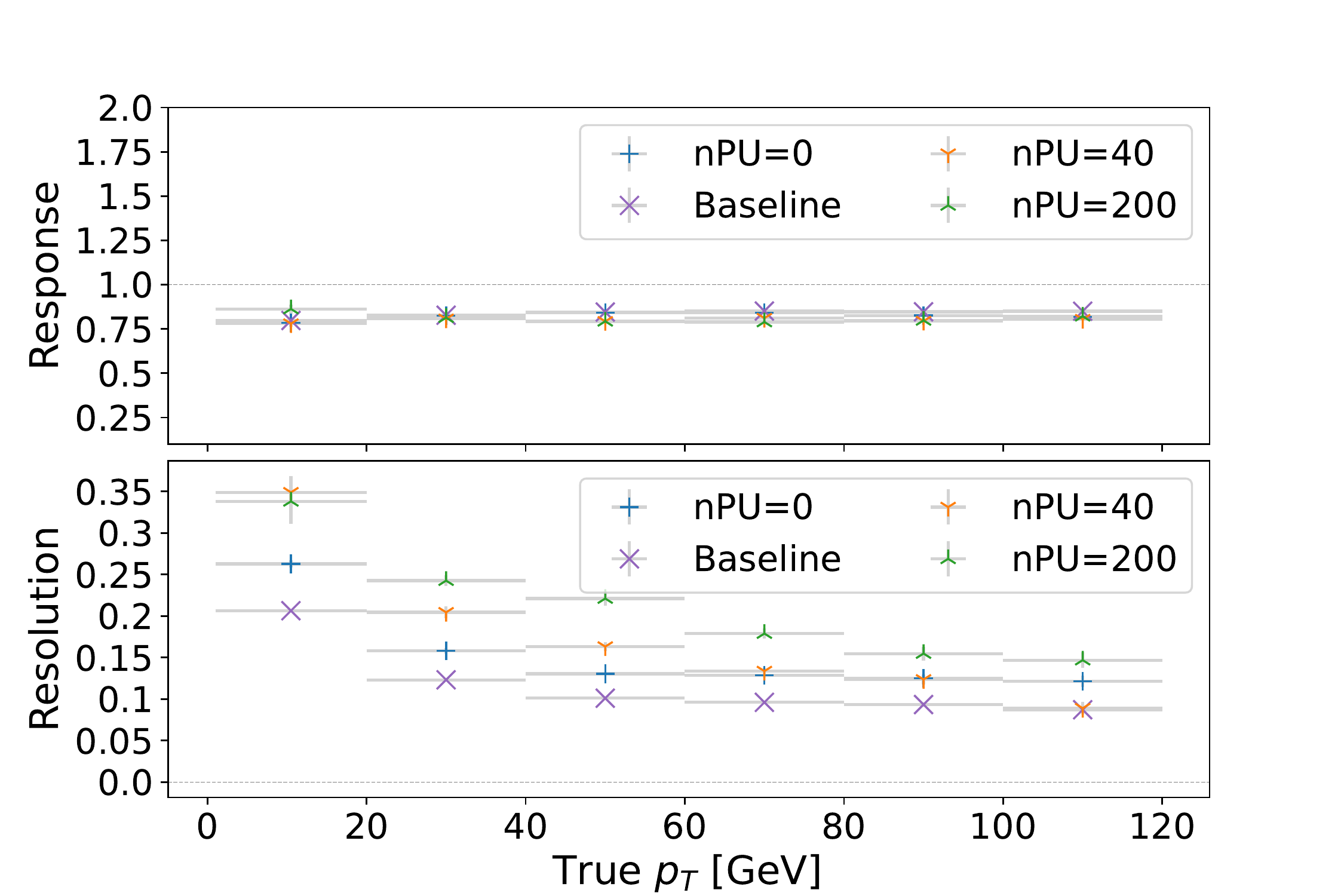}\label{fig:had_res}} \\
   \subfigure[]{\includegraphics[width=.55\textwidth]{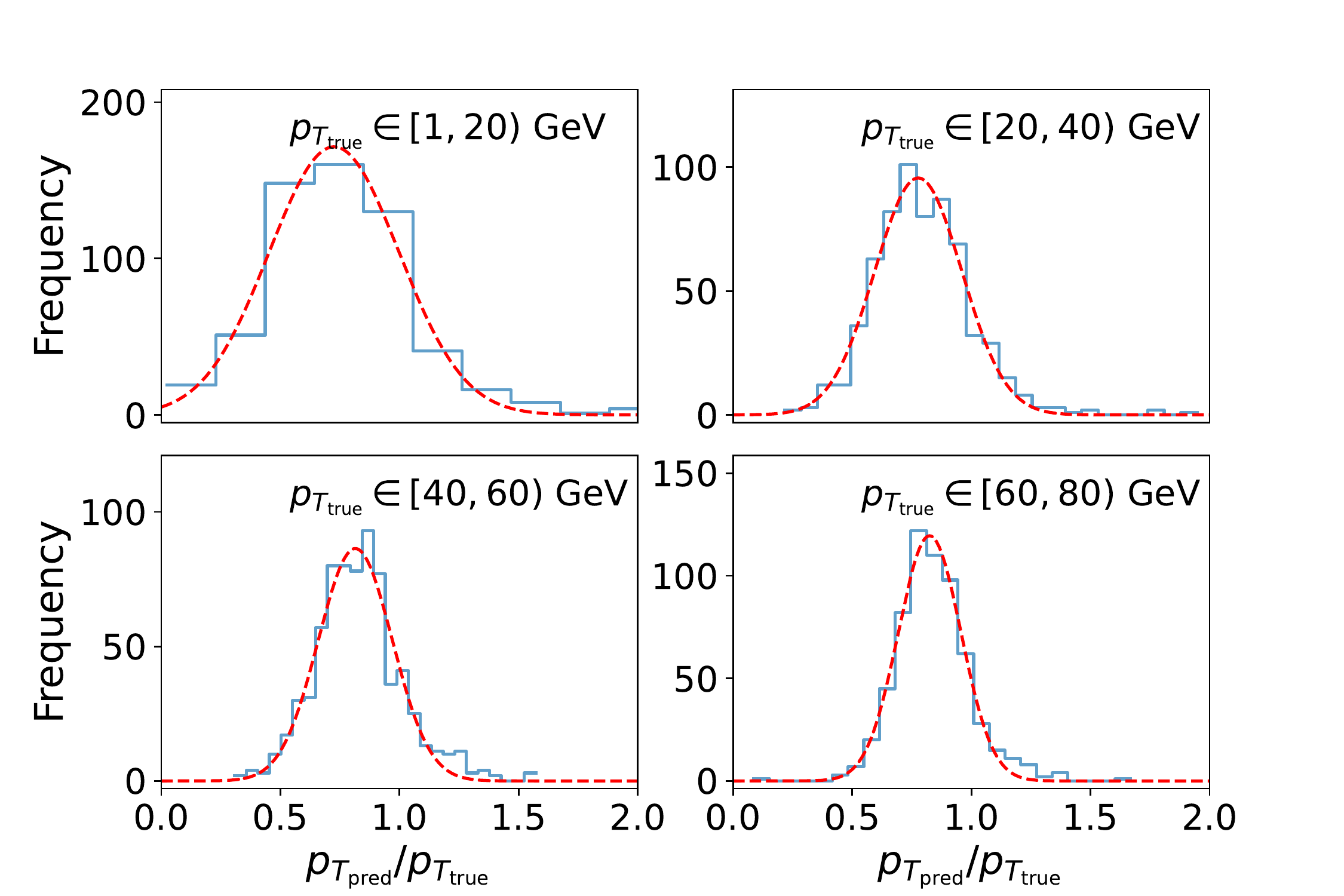}\label{fig:had_fit}}
    \caption{Reconstruction performance of hadrons ($\pi^+$) in different pileup environments. Fig.~\ref{fig:had_eff}: Efficiency as a function of the true $p_T$. Fig.~\ref{fig:had_res}: Mean response (top) and resolution (bottom) as functions of the true $p_T$. The response and resolution are computed as the mean and mean-corrected standard deviation of the Gaussian fit to the ${{p_T}_{\mathrm{pred}}}/{{p_T}_{\mathrm{true}}}$ distribution in individual $p_T$ bins. Fig.~\ref{fig:had_fit}: The distribution of response in different $p_T$ ranges corresponding to the first four bins in Fig.~\ref{fig:had_eff} and Fig.~\ref{fig:had_res} in 0 pileup environment.}
    \label{fig:had_results}
\end{figure*}

First, we match the probe shower to one of the predicted showers by applying a hit-based matching procedure that we already introduced in Ref.~\cite{qasim2021multi}. The procedure calculates energy weighted hit-intersection over hit-union score (EIOU) of a reconstructed cluster ${p}$ and a truth shower $\hat{t}$. The predicted shower that results in the highest overlap is taken as the matched shower ($\hat{p}$):
\begin{equation}
    \label{eq:matching_1}
    \hat{p} = \underset{p\in P}{\mathrm{argmax}}(\mathrm{EIOU}(\hat{t}, p)) \text{.}
\end{equation}
We apply a lower threshold of 0.5 to the EIOU score to study reconstruction efficiency which is shown in Fig.~\ref{fig:em_eff} and Fig.~\ref{fig:had_eff} for electromagnetic and hadronic particles respectively. 

The efficiency rises steeply with the increase in $p_T$ in both electromagnetic and hadronic cases. In 0 pileup, the efficiency reaches a plateau of almost one at $p_T > 1$~GeV for electromagnetic particles while it remains slightly lower for the hadronic particles with $p_T < 15$~GeV. As expected, because of the dense environment, the performance drops as the pileup is increased. In 40 pileup, the reconstruction efficiency of the electromagnetic particles deteriorates to around $80\%$ at $1$-$15$~GeV with a significant drop for $p_T < 1$~GeV. For the hadronic particles the reconstruction efficiency drops to around $70\%$ at $5$-$20$~GeV.

The efficiency deterioration occurs when the neural network oversplits the showers and these split showers fail to satisfy the matching criterion. Unlike electromagnetic showers, the tendency to oversplit is inherent to the nature of hadrons which is why the hadronic efficiency drops also at high $p_T$ when pileup is added. Therefore, to study the oversplits, we use Energy-Weighted Intersection Over Minimum (EIOM), defined below:

\[ \mathrm{EIOM}(t,p) =  \frac{\sum_{h \in (H_t \cap H_p)} e_h}{\mathrm{min}(\sum_{h \in H_t} e_h,\sum_{h \in H_p} e_h)}\]

\begin{figure}[t!]
    \centering
    \includegraphics[width=0.5\textwidth]{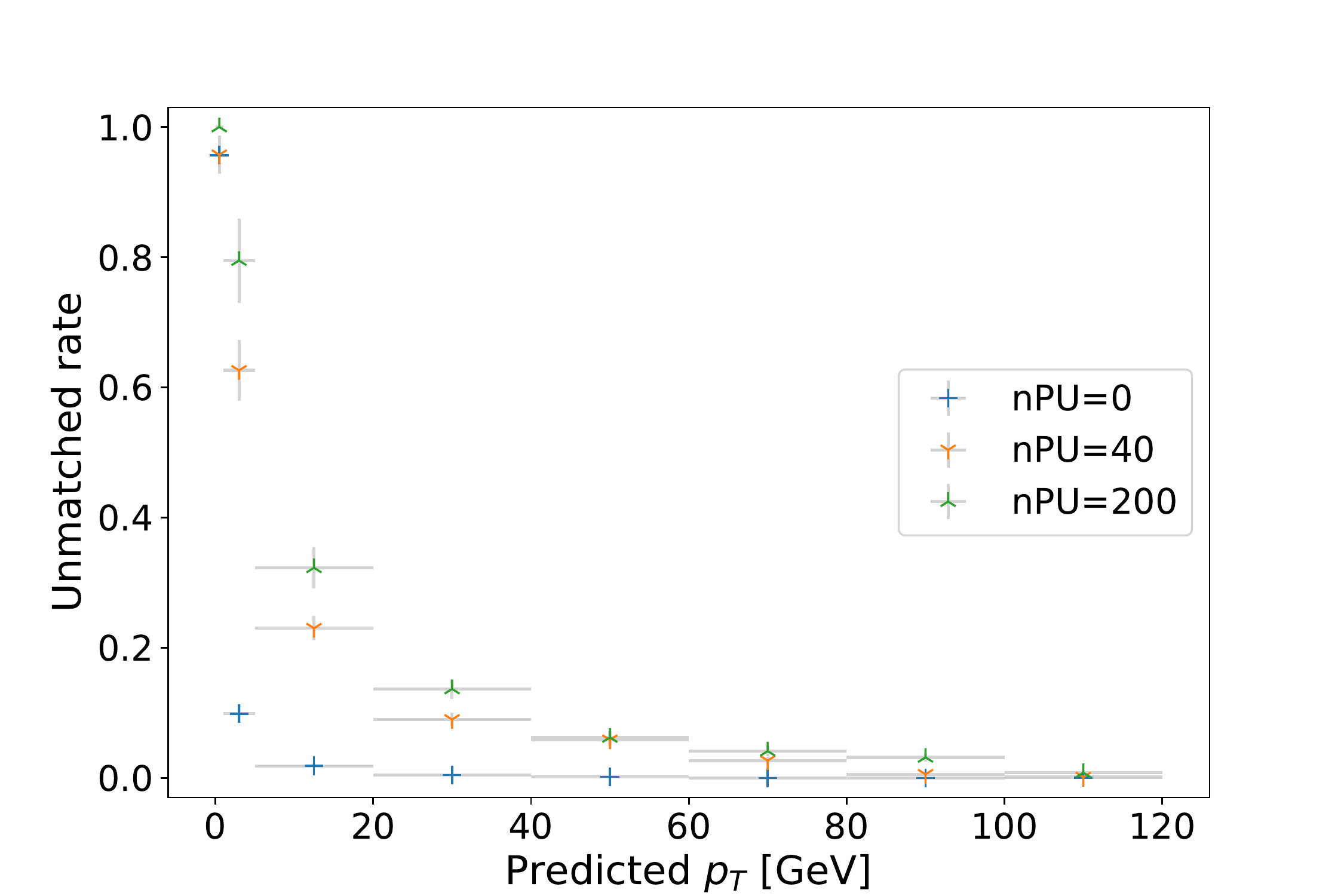}
    \caption{Unmatched rate as a function of predicted $p_T$.} 
    \label{fig:unmatched}%
\end{figure}

Unmatched showers are all the predicted clusters with EIOM $> 0.9$ with the truth-level probe particle but with EIOU less than $0.5$ and these are shown in Fig.~\ref{fig:unmatched}. The unmatched rate decreases steeply with the predicted $p_T$. This indicates that low $p_T$ clusters are split off from higher-$p_T$ showers while most of the energy is reconstructed properly. We note, that by adding tracking information and employing a suitable particle flow algorithm, these splits could be re-merged, increasing the efficiency. In addition to oversplitting, it could be possible that the neural network creates fake showers from the noise hits only. However, we observed that the fake rate is close to zero for $p_T$ above $1$~GeV. Additionally, only $0.5\%$ of the total noise energy is assigned to a predicted cluster on average.



For the matched true showers, we compare the true $p_T$ and, for the corresponding predicted shower, the regressed $p_T$. As a baseline, we define a truth-assisted $p_T$ reconstruction that considers the true deposited energy as reconstructed energy which is also then compared to the true shower $p_T$.
In Fig.~\ref{fig:em_res} and Fig.~\ref{fig:had_res}, we show the $p_T$ regression performance scaling with the truth $p_T$ for electromagnetic and hadronic particles, respectively. To compute the response and resolution, we fit the distribution of the $p_T$ response, ${p_{T}}_{\mathrm{pred}}/{p_{T}}_{\mathrm{true}}$, with a Gaussian function independently in each $p_T$ bin. Fig.~\ref{fig:em_fit} and Fig.~\ref{fig:had_fit} show the $p_T$ response distributions and the Gaussian functions for the first four bins. Mean ($\mu$) and mean-corrected standard deviation ($\sigma/\mu$) of the fitted Gaussian functions are taken as the response and resolution. The response can be corrected a posteriori, although it serves as an important metric for the algorithm's behavior, while the resolution directly reflects the $p_T$ reconstruction performance.


Due to the nature of hadronic showers and because the cell energies are calibrated on electromagnetic showers, the baseline response for charged pions is below one while it is compatible with one for electromagnetic showers, in particular at high energies. The reconstructed $p_T$ response provided by the network only differs mildly from the baseline response. The difference to unity response increases by about a factor of two for charged pions if the energy correction factor is not applied (not shown), indicating that the network is capable of distinguishing different shower types.

As expected, the resolution improves with true $p_T$ and degrades with increase in pileup. Also, here, even the reconstructed hadronic shower resolution is close to the baseline reconstruction and converges to $\sim 15\%$ above  $60$ GeV, even in high pileup. The reconstructed energy resolution in 0 pileup for electromagnetic showers is almost indistinguishable from the baseline and, therefore, close to the detector limitations. In 40 pileup, the electromagnetic resolution deviates from the baseline only at low $p_T$ but approximates the baseline at high $p_T$ and in 200 pileup, it deviates slightly more.


The neural network was trained on events with hundreds of true showers as shown in Table~\ref{tab:event_complexity_x} however it offers an excellent generalization performance on a vastly different datasets, especially where only one particle is present (0 pileup environment). The fact that we do not observe the creation of fakes demonstrates that the neural network has correctly learned to cluster using only local information. This increases confidence in the extrapolation capabilities of the network and training method beyond training conditions in general.

\subsection{Jet reconstruction performance}
Jets and their substructure are an integral ingredient for the analysis of particle collisions. In particular in the forward region, well-resolved individual jet constituents are crucial for a successful pileup removal and for identifying e.g. quark jets over gluon jets in vector-boson-scattering or fusion processes. As modern jet clustering algorithms are infrared and collinear safe, jets also offer a way to gauge the performance of the calorimeter clustering algorithm without strong dependencies of subtleties in the definition of the single-particle truth. Moreover, oversplitting and overmerging due to the reconstruction process have less impact on the cumulative jet quantities. 
Generally, a pileup removal algorithm is applied before jet clustering to remove contributions from particles not associated to the primary collision. As discussed in section~\ref{sec:data}, we generate $q\overline{q} \to t\overline{t}$ events on top of either 40 or 200 minimum bias events, as these events provide significant hadronic activity in the forward region. After the single-particle reconstruction is performed, a pileup removal algorithm is simulated aided by truth information:
We remove all showers that originate from pileup interactions unless they share more than 10\% of their energy-weighted hits with a reconstructed non-pileup shower. All remaining reconstructed showers are considered for clustering the reconstructed jets. To form the truth jets, we only consider truth showers that stem from the non-pileup interaction. We also define a baseline reconstruction algorithm based on the true deposited energy of the incident non-pileup particles.
Jets are then clustered using the anti-kt algorithm \cite{cacciari2008anti}  with a distance parameter of $R=0.4$. Reconstructed and truth jets are matched based on a  $\Delta R = \sqrt{\Delta \eta^2 + \Delta \phi^2}$  matching. Among all jets with $|\Delta p_T| / {p_T}_\mathrm{true} < 0.5$ and $\Delta R < 0.3$, we select the best match by minimum $\Delta R$.

\begin{figure*}[ht]
   \centering
   \subfigure[]{\includegraphics[width=.45\textwidth]{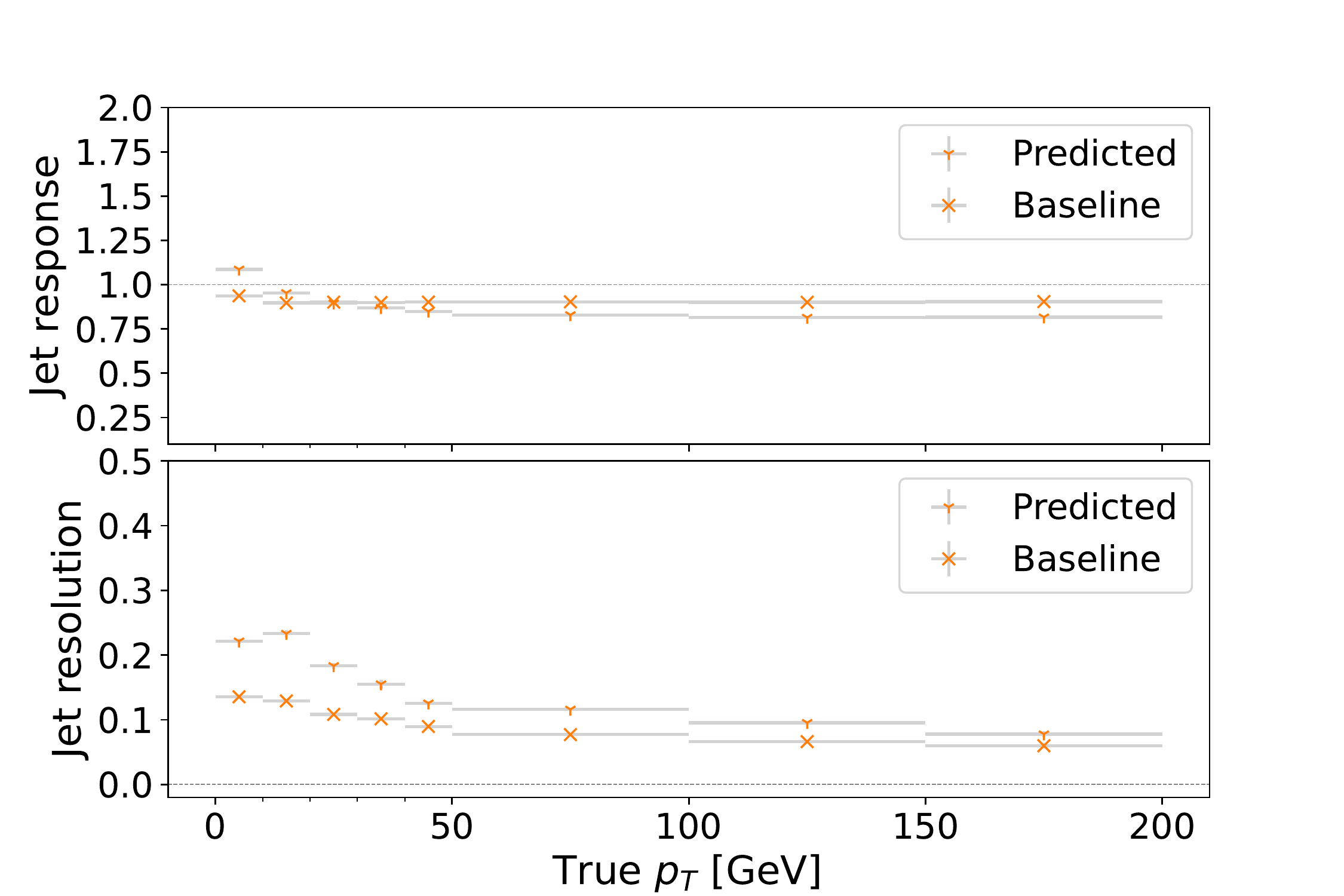}\label{fig:jet_40}}\quad
   \subfigure[]{\includegraphics[width=.45\textwidth]{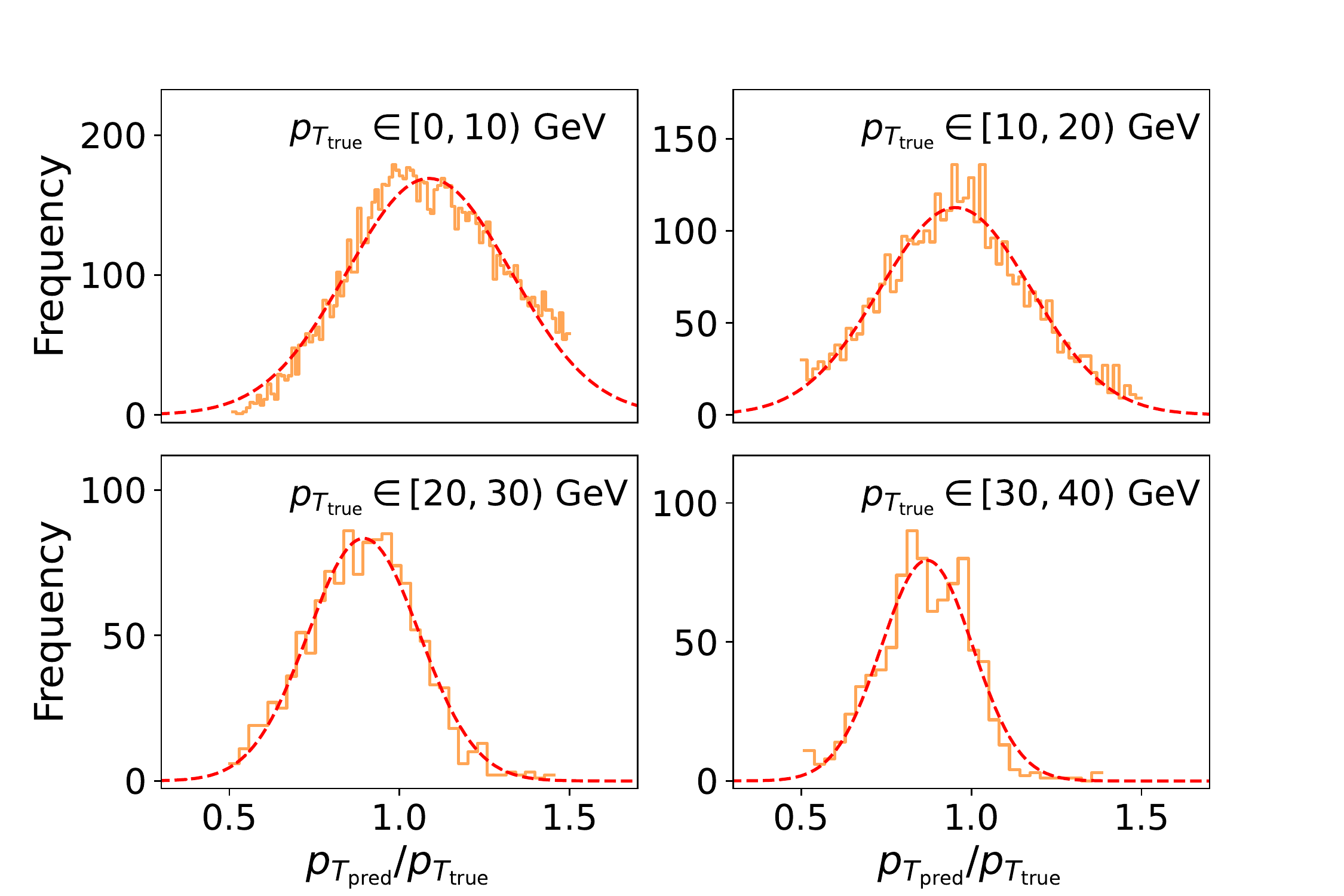}\label{fig:jet_40_fit}} \\
   \subfigure[]{\includegraphics[width=.45\textwidth]{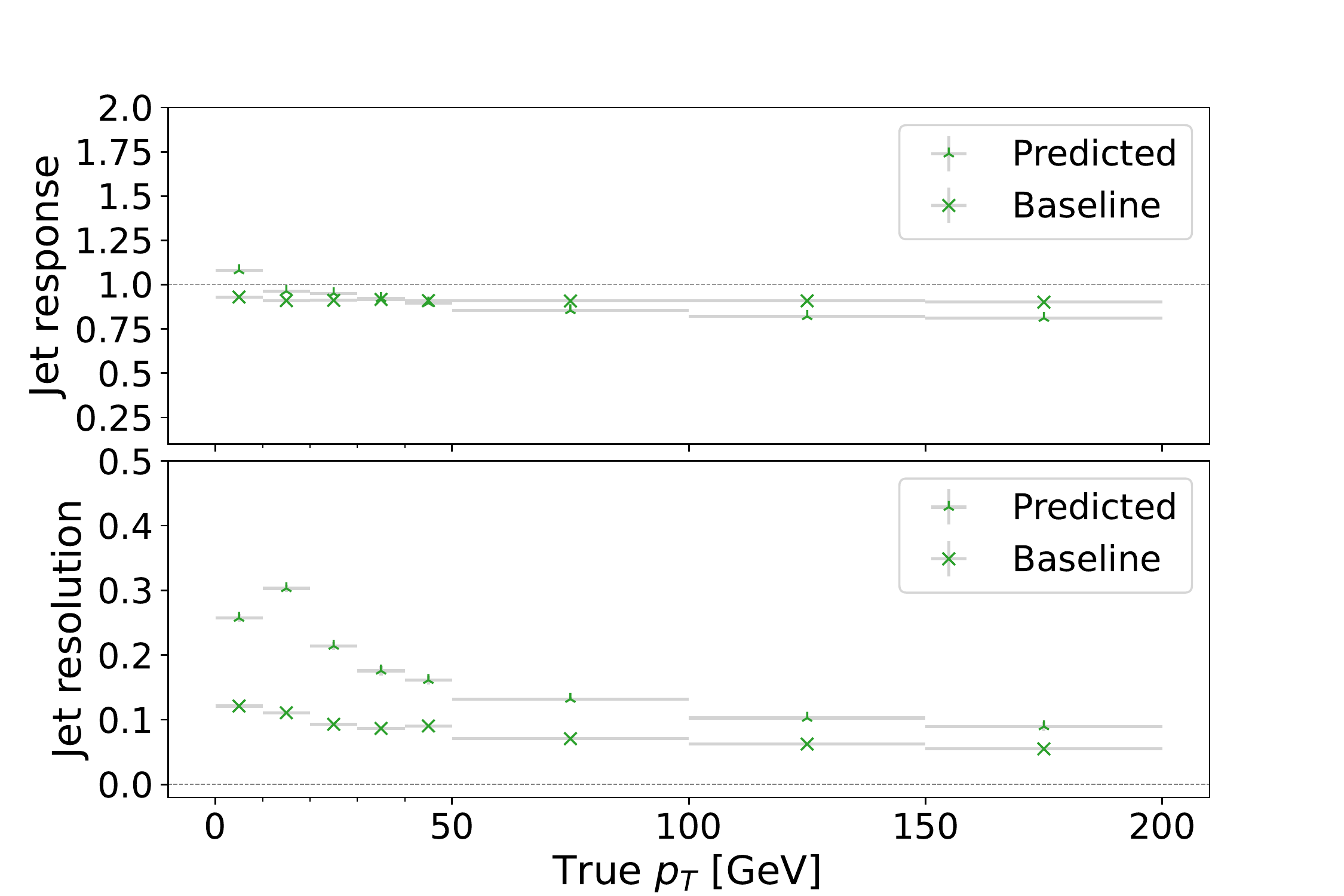}\label{fig:jet_200}}\quad
   \subfigure[]{\includegraphics[width=.45\textwidth]{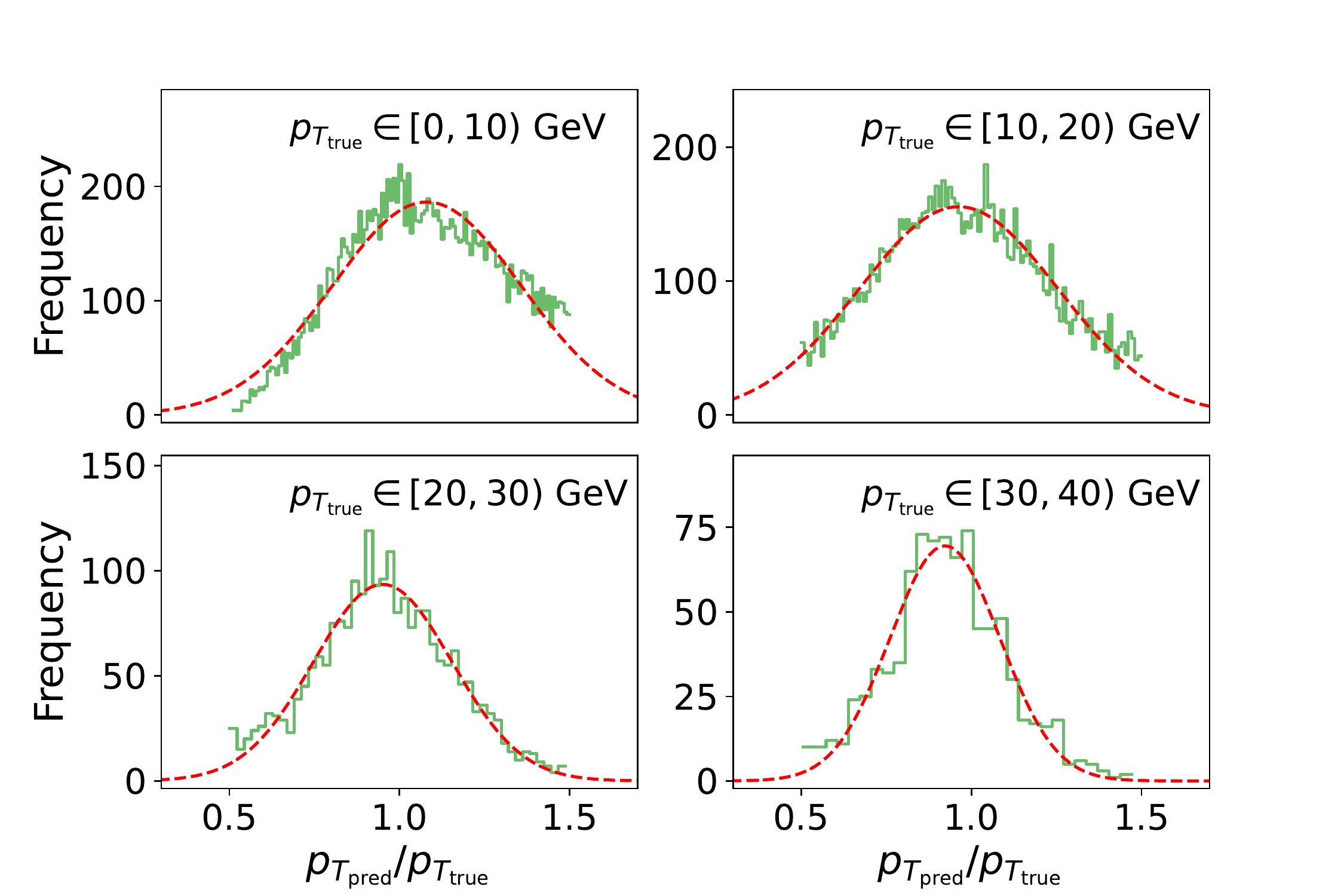}\label{fig:jet_200_fit}}
   \caption{Jet reconstruction performance. Fig.~\ref{fig:jet_40}: Mean jet response (top) and resolution (bottom) in 40 pileup as functions of the true $p_T$. Fig.~\ref{fig:jet_40_fit}: Response distributions and the fitted Gaussian functions in different $p_T$ ranges corresponding to Fig.~\ref{fig:jet_40}. Fig.~\ref{fig:jet_200}: Mean jet response (top) and resolution (bottom) in 200 pileup as functions of the true $p_T$. Fig.~\ref{fig:jet_200_fit}: Response distributions and the fitted Gaussian functions in different $p_T$ ranges corresponding to Fig.~\ref{fig:jet_200}. In both 40 and 200 pileup, the response and resolution are computed as the mean and mean-corrected standard deviation of the Gaussian fit to the ${{p_T}_{\mathrm{pred}}}/{{p_T}_{\mathrm{true}}}$ distribution in individual $p_T$ bins. We note that the better resolution in the first $p_T$ bin with the respect to the second bin is an artifact of the biased Gaussian fit due to the asymmetry of the response distribution, as shown in the top-left subplots of Fig.~\ref{fig:jet_40_fit} and Fig.~\ref{fig:jet_200_fit}.}
   \label{fig:jet_results}
\end{figure*}

Following the procedure performed for individual particle reconstruction, we fit Gaussian functions to the $p_T$ response distribution in each $p_T$ bin. The mean ($\mu$) and the mean-corrected standard deviation ($\sigma/\mu$) of the Gaussian function are taken as jet response and resolution. The distributions and the fitted functions are shown in Fig.~\ref{fig:jet_40}, and the response and resolution are shown in Fig.~\ref{fig:jet_40} and Fig.~\ref{fig:jet_200} for 40 and 200 pileup, respectively.

The response falling below one is a direct consequence of the single-particle responses shown in Fig.~\ref{fig:had_res}. At higher energies, the resolution starts to approximate the baseline. At lower energies, the presence of pileup degrades the performance slightly, however much less than in the case of single particles. As jets are less affected by truth matching and splitting effects, the assumption that single-particle performance depends on the matching procedure and the splitting of showers is verified. As shown in Fig.~\ref{fig:jet_200}, the performance degrades in 200 pileup compared to 40 pileup.


\section{Conclusion and further research}
\label{sec:conclusion}


In this paper, we presented the first demonstration that end-to-end reconstruction in high granularity calorimeters using graph neural networks is a feasible method for event reconstruction in very dense particle physics collisions. We perform this task in a single-step approach, in which the particle hits in the detector are taken as inputs and clustered showers, including their corrected energy, are output with no intermediate steps. Our model has been built using the GravNet graph neural network and the object condensation approach. We have evaluated the performance of our model on both single-particle events and physics events with pileup, and demonstrated promising performance of energy resolution and response for single particles, as well as for clustered jets. The method shows excellent generalisation properties from single-particle events to dense jets in 200 pileup. 

The proposed method also provides computational advantages, allowing one to exploit GPU acceleration at inference hence reducing the average inference time by at least two orders of magnitude with respect to high-pileup projections of the currently employed algorithms designed for CPUs. 

Currently, our model does not perform particle identification on the cluster object, which will be further studied in future work. In addition, measurements from additional detector subsystems, such as timing or tracking information, could be included to extend the one-shot reconstruction to a full particle flow prediction. The results presented here pave the way for such extensions and represent an initial demonstration that one-shot inference is an effective and efficient means for reconstruction in dense particle physics collisions.

\section*{Acknowledgments}
This project has received funding from the European Research Council (ERC) under the European Union's Horizon 2020 research and innovation program (Grant Agreement No. 772369).

We thank Ian Fisk and the Flatiron Institute of the Simons Foundation for providing access to the GPU cluster used to perform these studies.


\bibliographystyle{unsrt}  
\bibliography{main}  

\appendix

\begin{figure*}[t!]
    \centering
    
   \includegraphics[width=.45\textwidth]{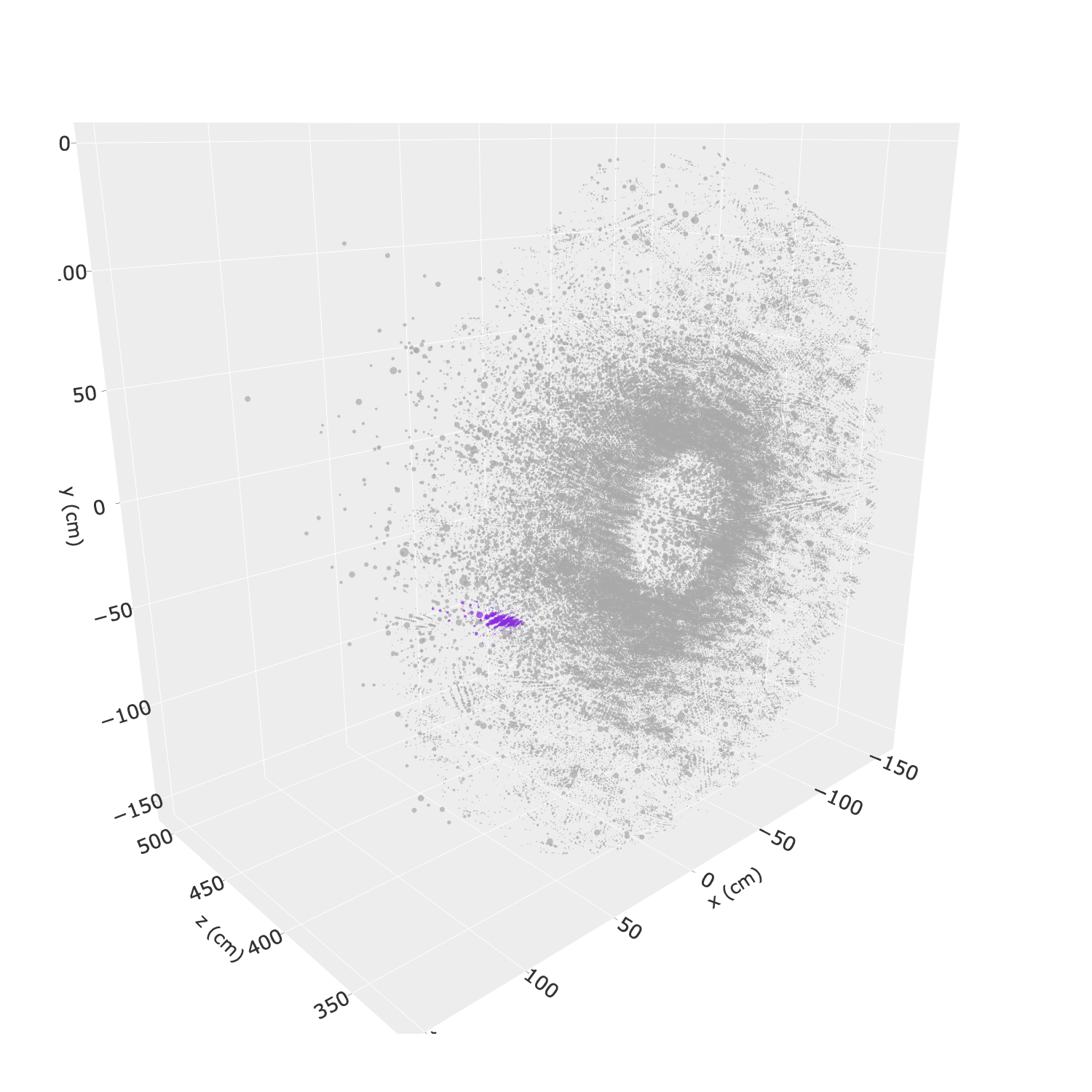}
   \includegraphics[width=.45\textwidth]{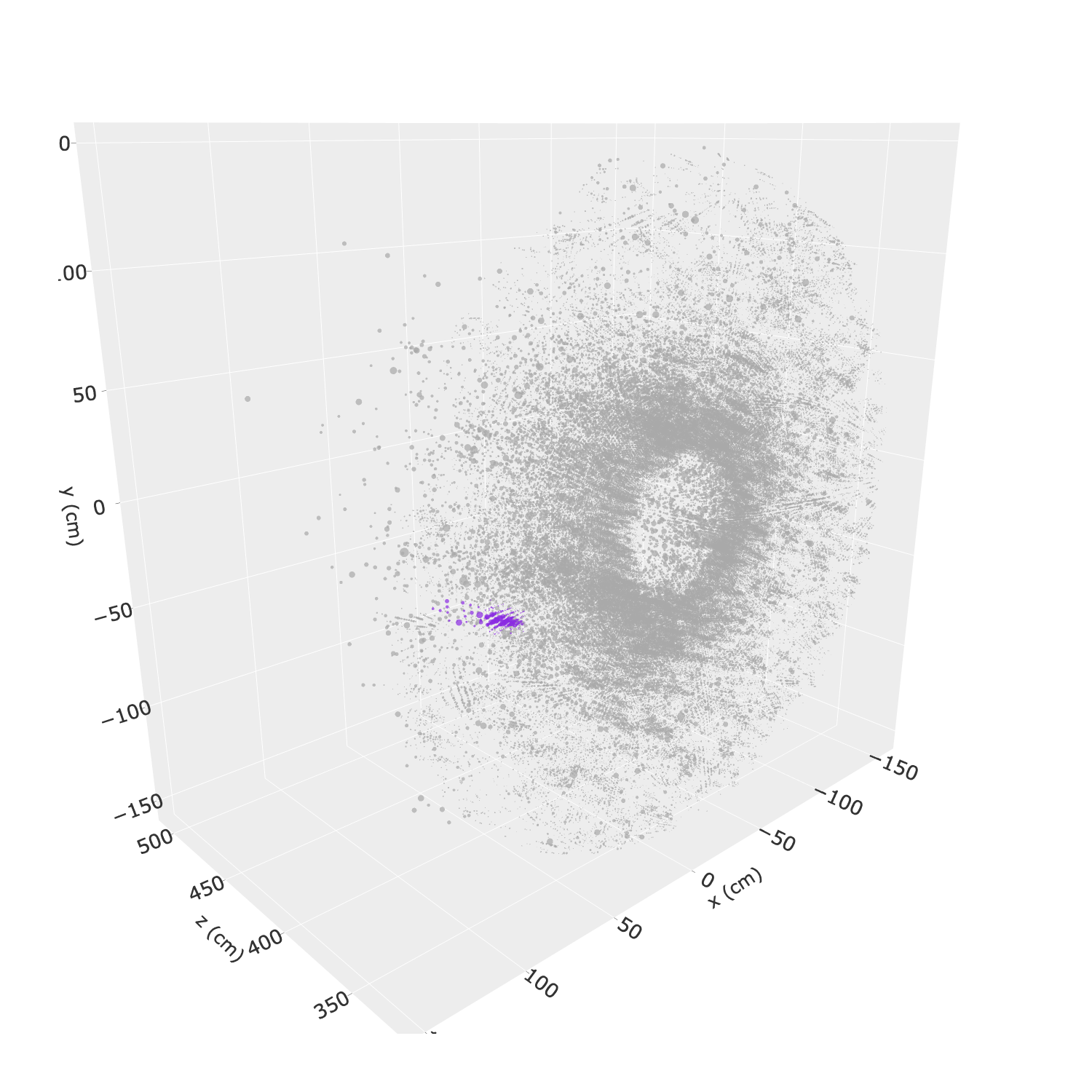} \\
  \includegraphics[width=.45\textwidth]{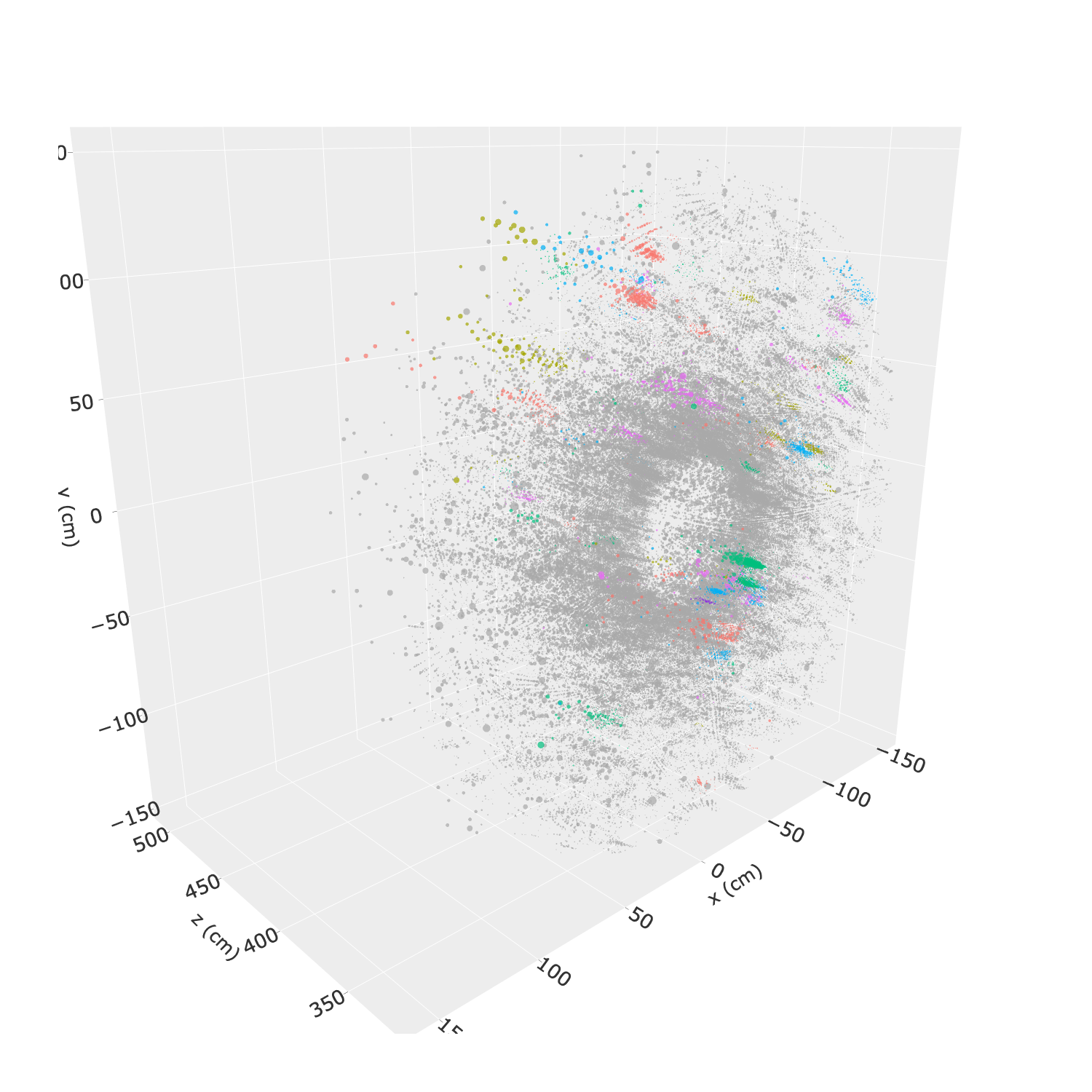}
\includegraphics[width=.45\textwidth]{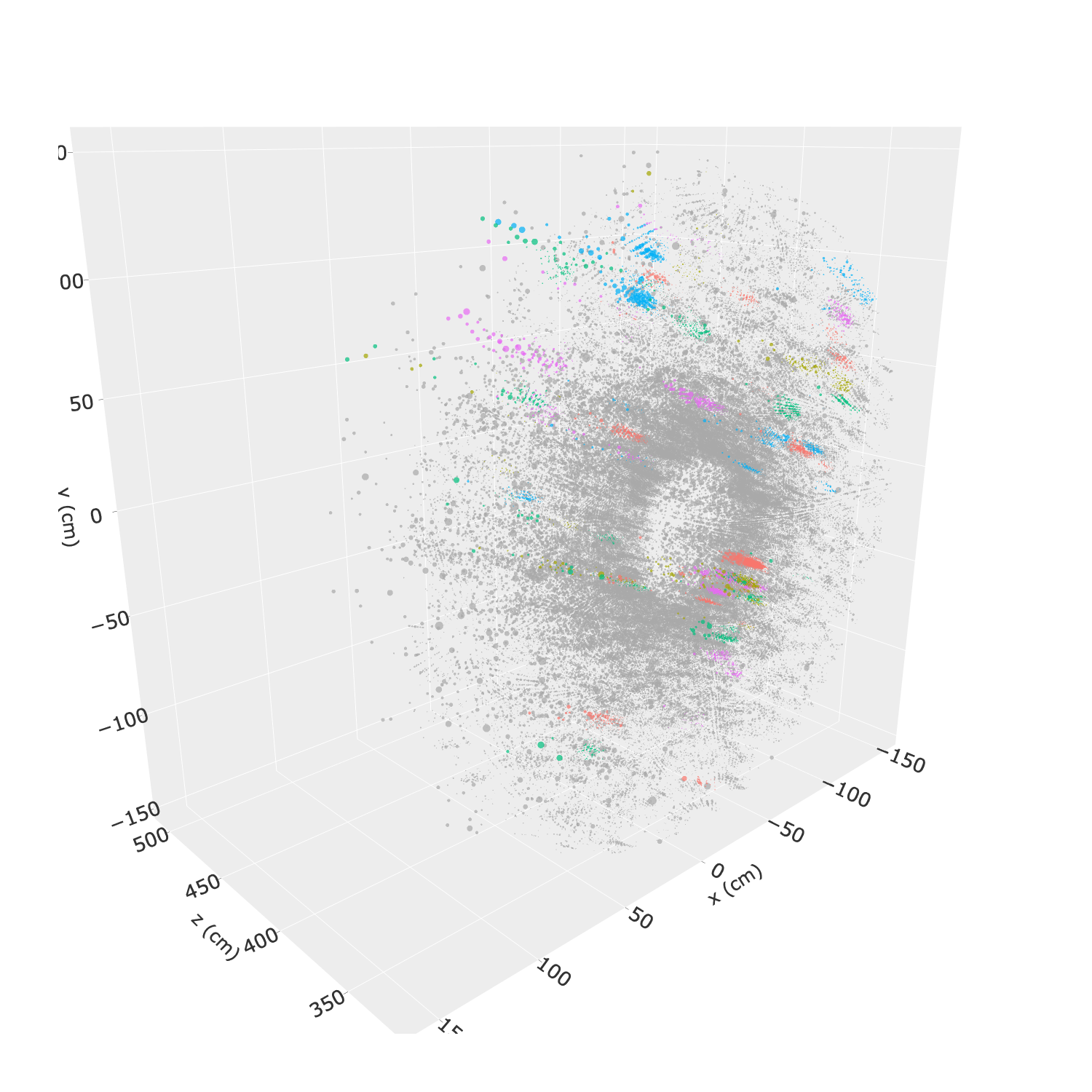} \\
    \caption{True versus predicted cluster examples in 200 pileup. In the top two figures, a single particle is shot into the calorimeter, where the left and right figures show the true and the corresponding matched predicted cluster, respectively. The bottom row shows particles originating from $q\overline{q} \to t\overline{t}$ collision in colors while the grey is 200 pileup. The predicted clusters (right) are matched to the true clusters (left). Jet reconstruction performance is studied on these true and matched clusters while the pileup is ignored.
    \label{fig:event_reco_examples}}
\end{figure*}

\printnomenclature

\end{document}